\documentclass[11pt,a4paper]{article}

\pdfoutput=1
\usepackage{jheppub}

\usepackage{color}
\usepackage{graphicx}
\usepackage{wrapfig,enumerate,slashed}

\usepackage{footmisc}
\usepackage{amsmath}
\usepackage{wasysym} 
\usepackage{graphicx}
\usepackage{color}
\usepackage{comment}
\usepackage{epstopdf}
\usepackage{caption}
\usepackage{subcaption}

\DeclareMathAlphabet{\mathbold}{U}{zeur}{b}{n}

\usepackage[labelfont=bf,font={sl,small}]{caption}

\renewcommand\[{\left[}
\renewcommand\]{\right]}

\def\beq{\begin{equation}}
\def\eeq{\end{equation}}
\def\[{\begin{equation}}
\def\]{\end{equation}}

\begin{document}
\numberwithin{equation}{section}

\title{The Higgsploding Universe}

\author[a]{Valentin V. Khoze}
\author[a]{and Michael Spannowsky}

\affiliation[a]{Institute for Particle Physics Phenomenology, Department of Physics, Durham University, DH1 3LE, UK}

\emailAdd{valya.khoze@durham.ac.uk}
\emailAdd{michael.spannowsky@durham.ac.uk}

\abstract{
Higgsplosion is a dynamical mechanism that introduces an exponential suppression of quantum fluctuations beyond the Higgsplosion energy scale $E_*$ and further guarantees perturbative unitarity in multi-Higgs production processes. By calculating the Higgsplosion scale for spin 0, 1/2, 1 and 2 particles at leading order, we argue that Higgsplosion regulates all n-point functions, thereby embedding the Standard Model of particle physics and its extensions into an asymptotically safe theory. There are no Landau poles and the
Higgs self-coupling stays positive. Asymptotic safety is of particular interest for theories of particle physics that include quantum gravity. 
We argue that in a Hippsloding theory one cannot probe shorter and shorter length scales by increasing the energy of the
collision beyond the Higgsplosion energy and there is a minimal length set by $r_* \sim 1/E_*$ that can be probed.
We further show that Higgsplosion in consistent and not in conflict with models of inflation and the existence of axions.
There is also a possibility of testing Higgsplosion experimentally at future high energy experiments.
}


\preprint{IPPP/17/53}

\maketitle



\section{Introduction}
\label{sec:intro}
\medskip
Higgsplosion and Higgspersion \cite{Khoze:2017tjt} are two intertwined mechanisms which aim to explain why physical scattering processes involving many scalars are not bound to violate unitarity, while solving at the same time the Hierarchy problem of the Higgs boson. 

It was shown in several independent calculations at leading order
\cite{Brown:1992ay,Argyres:1992np,Voloshin:1992rr,Libanov:1994ug,Khoze:2014kka}, 
one-loop resummed \cite{Voloshin:1992nu,Smith:1992rq,Libanov:1994ug} or using a semiclassical approach 
\cite{Son:1995wz,Gorsky:1993ix,Khoze:2017ifq}, that the rate for a transition $h^* \to n \times h$ grows factorially with $n$.
Based on these results it was expected that either the cross section of physical processes, e.g. in proton collisions of $pp \to n \times h$, 
has to grow as well, thereby violating perturbative unitarity in tree-level processes above certain critical values of collision energy and 
multiplicity and pointing to a validity limit of the standard model of particle physics, or that a non-trivial energy-dependent formfactor
had to emerge in a strongly coupled perturbation theory or even non-perturbatively.
A second short-coming of the standard model, identified early on, is that the Higgs boson is known to receive quadratic contributions from quantum corrections $m_h^2 \simeq m_0^2 + \delta m_{\mathrm{new}}^2$. Thus, in order to obtain the physical mass $m_h \simeq 125$ GeV, the bare parameter of the theory $m_0$ has to be increasingly precisely tuned, depending on the hierarchy between the electroweak and new physics scale. With the discovery of an elementary scalar particle \cite{Aad:2012tfa,Chatrchyan:2012xdj}, both of these problems cannot be ignored if the standard model is not to be fundamentally modified at energy scales of $\mathcal{O}(100)$ TeV, i.e. energy scales possibly within reach of a future hadron collider \cite{Contino:2016spe}.

In \cite{Khoze:2017tjt} we argued that the factorial growth of the amplitude $h^* \to n \times h$ or $X \to n \times h$ beyond a Higgsplosion scale $E_*$ can be directly related to the exponential growth of the  imaginary part of the of the width of $h$ and X using the optical theorem, thus avoiding violation of perturbative unitarity in the production of many Higgs bosons and suppressing high-scale contributions of $X$ to the mass of the Higgs boson. Hence, the Higgs boson can provide a self-consistent solutions to the problems it introduced to the standard model.

Assuming Higgsplosion and Higgspersion were realised in nature, the question this paper aims to address is what are their phenomenological consequences and in how far is it possible to address and solve other fundamental questions in high-energy physics, beyond 
providing a solution \cite{Khoze:2017tjt} to the Hierarchy or fine-tuning problem of the Higgs boson in the standard model.
We will describe the scope of Higgsplosion, pointing out that not only the Higgs boson and heavy resonances that can decay into many Higgs bosons higgsplode at a scale $E_*$, but the same fate can affect all particles, in particular all standard model particles. We further show that Higgsplosion can provide a self-consistent picture of nature up to very high energy scales and a rich phenomenology with direct implications to high-scale physics.
\vspace{0.2cm}

We will discuss the quantum field theoretical consequences of Higgsplosion and Higgspersion in Sec.~\ref{sec:qft}. By explicitly calculating the effect of Higgsplosion based on leading-order amplitudes for the Higgs boson (spin-0), the top quark (spin-1/2), a vector boson (spin-1) and the graviton (spin-2) we show that all standard model particles higgsplode and 
when this happens,  all n-point functions are rendered finite. Further, not only high loop momenta are suppressed but in general all processes with propagators with high $p^2$, leaving in collisions at very high energies only $t$-channel processes while rendering $s$-channel processes negligible. We also argue that in the higgsploding theories one cannot probe distances shorter than a certain minimal 
scale $r_*$ that is set by the inverse energy of Higgsplosion $E_*$. Above this energy, the theory screens the processes that attempt to 
probe shorter length scales. This is the effect of Higgspersion. In parallel we argue that higgsploding theories are asymptotically safe:
all coupling constants reach finite values in the UV regime above  $E_*$, there are no Landau poles and the Higgs self-coupling
$\lambda$ remains positive.
In Sec.~\ref{sec:pheno} we outline the consequences of Sec.~\ref{sec:qft} on some aspects of particle phenomenology. We discuss the running of the coupling constants and  show that Higgsplosion can provide an embedding of the standard model into an asymptotically-safe framework. We also explain that Higgsplosion is consistent and not in conflict neither with inflation nor with the existence axions.
In Sec.~\ref{sec:concl} we offer a brief summary and conclusions.

\medskip
\section{Quantum Field Theory in presence of Higgsplosion}
\label{sec:qft}
\bigskip

\subsection{The interplay between Higgsplosion and Higgspersion in physical processes}
\label{sec:2.1}

Higgsplosion \cite{Khoze:2017tjt} denotes the rapid decay of a heavy or highly energetic resonance $X$ into multiple Higgs bosons. 
The initial state $X$ can in the simplest case be the Higgs boson itself, which we denote as $h^*$ to emphasise that it is a highly virtual
state with the momentum $p^2 \gg M_h^2$, or equivalently it can be another Standard Model degree of freedom with the energy 
sufficiently high,
$\sqrt{p_2}= \sqrt{s} >n M_h$, so that its decay into final states involving a large number $n \gtrsim 100$ of Higgs bosons is kinematically
possible. Importantly, the state $X$ can also represent a very heavy new physics state with the mass $M_X \gg n M_h$, which can then
decay into multiple Higgs bosons already at momentum scales $\sqrt{p_2}= \sqrt{s} >n M_h$ that are below its mass shell $\ll M_X$.
The latter case would correspond to a heavy state which potentially decays much before it can be realised as an on-shell particle state,
and this set-up is important for addressing the Hierarchy problem by taming the radiative corrections from the virtual loops of $X$ contributing
to the Higgs mass parameter, see Ref.~\cite{Khoze:2017tjt} and section~\ref{sec:2.3} below.

In several independent calculations at leading order
\cite{Brown:1992ay,Argyres:1992np,Voloshin:1992rr,Libanov:1994ug,Khoze:2014kka}, 
one-loop resummed \cite{Voloshin:1992nu,Smith:1992rq,Libanov:1994ug} or using a semiclassical approach 
\cite{Son:1995wz,Gorsky:1993ix,Khoze:2017ifq}, 
it was shown that the rate for a transition $h^* \to n \times h$ grows factorially with $n$, 
where the state's total and partial widths are respectively given by
\[
\Gamma (p^2) \,=\, \sum_{n=2}^\infty \Gamma_n (p^2) \,\,, \qquad
\Gamma_n (p^2)\,=\,\frac{1}{2M_h} \int \frac{1}{n!} d\Pi_n |{\cal M}(h^* \to n\times h)|^2 \,. \label{eq:2.11}
\]
with $|{\cal M}|^2$ being the scattering amplitude squared which is integrated over the $n$-particle phase space $d\Pi_n$
including the bosonic symmetry factor $\frac{1}{n!} $, and $M$ is the mass.
With the recent calculation of \cite{Khoze:2017ifq} the parametric dependence of the Higgs boson's partial width on the number of final state Higgs bosons $n$, their average kinetic energy $\varepsilon = (\sqrt{p^2} - n M_h)/(n M_h)$ and the Higgs self coupling $\lambda$ can be expressed by 
\begin{equation}
\Gamma_n(p^2) \, \propto {\cal R}(\lambda; n,\varepsilon) = \exp \left[ \frac{\lambda n}{\lambda}\, \left( 
\log \frac{\lambda n}{4} \,+\, 3.02\, \sqrt{\frac{\lambda n}{4\pi}}\,-\,1\,+\,\frac{3}{2}\left(\log \frac{\varepsilon}{3\pi} +1 \right)
\, -\,\frac{25}{12}\,\varepsilon 
\right)\right] \,  .
\label{eq:Rn}
\end{equation}
Here we defined the dimensionless variable ${\cal R}(\lambda; n,\varepsilon)$ or for brevity ${\cal R}_n$ which is 
the rescaled $n$-particle decay rate ${\cal R}_n (p^2)=\Gamma_n(p^2)/M_h = -{\rm Im} \Sigma (p^2)/M_h^2$.

The expression \eqref{eq:Rn} was derived in the combined weak-coupling--large multiplicity limit, $\lambda \to 0$, $n\to \infty$,
in the regime where the final state particles are assumed to be non-relativistic, $\varepsilon \ll 1$, and the effective coupling 
parameter $\lambda n$ is large, $\lambda n \gg 1$. 
The characteristic exponential factor in \eqref{eq:Rn} has a semi-classical origin and it was argued in \cite{Libanov:1994ug,Son:1995wz}
that it is not affected by the choice of $X$ in the initial state in so far as $X$ is coupled to Higgs bosons.

Using the optical theorem, the decay width of $X$ can be directly related to the imaginary part of its self-energy, which results in a rapidly growing imaginary part of the denominator of its dressed propagator 
\[ \Delta_X (p)\,=\, 
 \frac{i }{p^2-M_X^2-i\, {\rm Im}\, \Sigma(p^2)} \,=\,
\frac{i }{p^2-M_X^2 +i \,M_X\, \Gamma(p^2)} \,.
\label{eq:pfin2}
\]

Due to the Higgsplosion effect 
it was believed for a long time that the cross-section of the physical processes, such as
the gluon fusion  $gg \to n\times h$ going through an intermediate virtual Higgs boson(s) produced in the $s$-channel,
$gg \to h^*\to n\times h$,
was bound to grow factorially with $n$ and would violate perturbative unitarity
at energy scales of $\mathcal{O}(100)$ TeV for $n \lambda \gg 1$,
(unless large quantum corrections at strong effective coupling $n \lambda \gg 1$ or non-perturbative physics would introduce a non-trivial suppression factor).

However, including the dressed propagators for intermediate $h^*$ in this process results in Higgspersion, i.e. a well-behaved cross section for arbitrary $n$ up to very high energies~\cite{Khoze:2017tjt}
\[ \sigma^{\Delta}_{gg \to n\times h} \, \sim \, 
y_t^{2}   
m_t^2  \log^4\left(\frac{m_t}{\sqrt{p^2}}\right) \,\times\, \frac{1}{p^4+M_h^4{\cal R}^2}
\,\times\, 
{\cal R}_n\,,
\label{eq:higgsper}
\]
and thus
\[
\sigma_{gg \to n\times h}\, \sim \, 
\begin{cases}
{\cal R} & :\,\,{\rm for}\,\,  {\cal R} \lesssim 1\\
1/{\cal R}\to 0 & :\,\,{\rm for}\,\,  {\cal R} \gg 1 \,\,{\rm at}\,\,p^2\to \infty\, .
\end{cases}
\label{eq:poly_ev_odd}
\]
Hence, by avoiding a breakdown of perturbative unitarity in multi-boson production, the theory can retain consistency and predictivity to much higher, technically even unlimited, energy scales.

\bigskip

Let us summarise what we mean by Higgsplosion:

\begin{enumerate}
\item Higgsplosion is triggered by the rapid exponential growth
of the $n$-particle decay rate $\Gamma_n \propto\, {\cal R}_n \to\, \infty $ with $n$ and $\sqrt{s}\sim n$,
as in \eqref{eq:Rn}. The Higgsplosion energy $E_*$ is where the rate changes from ${\cal R}_n \ll 1$ to ${\cal R}_n\gg 1$. It is a new dynamically generated non-perturbative scale in the theory,
$E_* = {\rm const}_*\, M/\lambda$, where $M_h$ and $\lambda$ are the mass and the coupling of the elementary scalars -- the Higgs bosons produced in the final state, and ${\rm const}_*$ is the calculable constant factor which is typically $ \gg 1$.
\item The initial building blocks of Higgsplosion are the Dyson-resummed propagaotrs for all degrees of freedom $X$ in the theory,
given by \eqref{eq:pfin2}.
They are obtained by summing up the geometric progression for 2-point functions 
over all self-energy insertions. The self-energy $\Sigma_X(p^2)$ 
is a function of the momentum of the propagator and it is represented by its large imaginary part due to Higgsplosion.\footnote{ Below, in 
section~\ref{sec:2.2} we will explain the rational why
${\rm Im} \Sigma_X(p^2)$ is expected to higgsplode, based on perturbative calculations at leading order.}
Because of the rapid increase of ${\rm Im} \Sigma_X(p^2)$ the propagators in \eqref{eq:pfin2} vanish for momenta exceeding the scale $p^2>E_*^2$. This is the result of highly energetic modes becoming unstable
against the multi-particle decays, and these modes at momentum scales above $E_*$ loose their particle interpretation.
The correct interpretation of the vanishing propagators is that the highly energetic field theoretical degrees of freedom become composite 
states made out of large numbers of relatively soft Higgs bosons and are no longer described by an individually propagating particle
degree of freedom. The dynamical scale of compositeness or classicalization, here an interaction and virtuality dependent experimentally unresolvable minimal length scale, is for each particle $i$ set by $E^i_*\sim 1/r_i$.

\item Technically, ${\rm Im} \Sigma_X(p^2)$ entering the expressions for Dyson-resummed propagators \eqref{eq:pfin2}, is obtained
from the decay rates $\Gamma_n$ using the optical theorem, and hence is determined in the physical domain $p^2=s>0$. 
However the underlying concept of compositeness should not depend on whether the momentum scale at which we probe the
particle is time-like or space-like. If this is true that a given degree of freedom becomes composite
in the physical domain of positive $p^2 > E_*^2$, the same should equally apply at space-like momenta, 
and only $|p^2| > E_*^2$ matters.
Hence in constructing the formalism for computing quantum (loop) corrections in Higgsploding theories we will describe the non-perturbative effect of compositeness caused by Higgsplosion in terms of the propagators 
\[ \Delta_X (p)\,=\, 
 \frac{i }{p^2-M_X^2} \, K(p^2/E_*^2) \,,
\label{eq:pfin2F}
\]
where $K(x)$ is the non-perturbative Higgsplosion (or more precisely Higgspersion) formactor,
\[
K(x)\, = \, 
\begin{cases}
\quad 1 & :\,\,{\rm for}\,\,  x <1\\
\quad 0 & :\,\,{\rm for}\,\, x> 1\, .
\end{cases}
\label{eq:Kx}
\]
In the physical domain and for $0< p^2 \le E_*^2$ there is little difference between the propagators \eqref{eq:pfin2} and
\eqref{eq:Kx}. Due to the sharp exponential rise of the decay rate ${\cal R}$ with energy in \eqref{eq:Rn}, both 
propagtors are sharply cut-off at or just above $p^2 = E_*^2$ and are vanishing above this value. At $p^2 < E_*^2$,
the multi-particle contribution to the decay rate is exponentially suppressed, to a very high accuracy it is zero, and both
expressions are correctly described by the bare propagators. (One can always add the additional non-Higsploding width effects to the denominators of both expressions if required.)

\item {The dressed propagators \eqref{eq:pfin2F} are then used as the {\it input} into the computation of $n$-point functions with $n\ge 3$.
All quantum contributions to the $n$-point vertices are obtained in our formalism by computing loop effects that involve these
dressed  propagators and ordinary renormalised vertices. This is done order by order in the loop expansion, where the internal lines in the loops
are given by the dressed propagators in the form \eqref{eq:pfin2F} treated as the input. Integrations over the loop momenta can 
now also be carried out in Euclidean space if desired.}
\item It then follows from this formalism that the theory is made UV-finite by Higgsplosion and the couplings are asymptotically safe,
as will be explained in section . There are no Landau poles and the Higgs self-coupling cannot not become negative and hence the electroweak vacuum is stable.
\item Finally it is also easy to check \cite{Khoze:2017tjt} using the same method that the real part of the self-energy is also UV-finite
and that even the finite fine-tuning of the quantum corrections to Higgs mass parameter from integrating out a very heavy $X$ state, 
$M_X \gg E_*$,  
 is reduced by many orders of magnitude from $\delta M_h^2 \sim M_X^2$ 
to $\delta M_h^2 \sim E_*^4/M_X^2$. This solves the Hierarchy problem.
\end{enumerate}

\bigskip

\subsection{Higgsplosion of the self-energy}
\label{sec:2.2}
\bigskip

The Higgsplosion effect becomes operative when the imaginary part of the self-energy $\Sigma_X (p^2)$
for a given field theoretical degree of freedom $X$ becomes exponentially large, i.e. when the external momentum 
 $p$ approaches the critical
energy scale $E_{*}$ of Higgsplosion. Specifically,
\[
{\rm Im} \,\Sigma_X (p^2) \, \sim\, {\cal R}_X\,, \quad{\rm where} \quad
\begin{cases}
{\cal R}_X \ll 1& :\,\,{\rm for}\,\,  p^2 < E_{*}^2\\
{\cal R}_X \gg 1& :\,\,{\rm for}\,\,  p^2 \gtrsim E_{*}^2\,.
\end{cases}
\label{Sigma_phi}
\]
The value of the Higgsplosion scale $E_{*}$ for $X$ in general depends on the nature and strength of interactions between $X$ 
and the Higgs bosons.

To study the implications and extend of Higgsplosion, we will now consider different choices for $X$, first by taking it to be the
Higgs boson itself; second another light degree of freedom (for example the top quark (spin-1/2), a vector boson (spin-1) or a graviton (spin-2)); and finally a
 heavy degree of freedom with the mass much greater than the electro-weak scale and unstable to decay into multiple Higgses.

\bigskip

\subsubsection{Higgsplosion in the self-energy of the Higgs}
\label{sec:2.2.1}

\begin{figure}[t!]
\begin{center}
\includegraphics[width=0.35\textwidth]{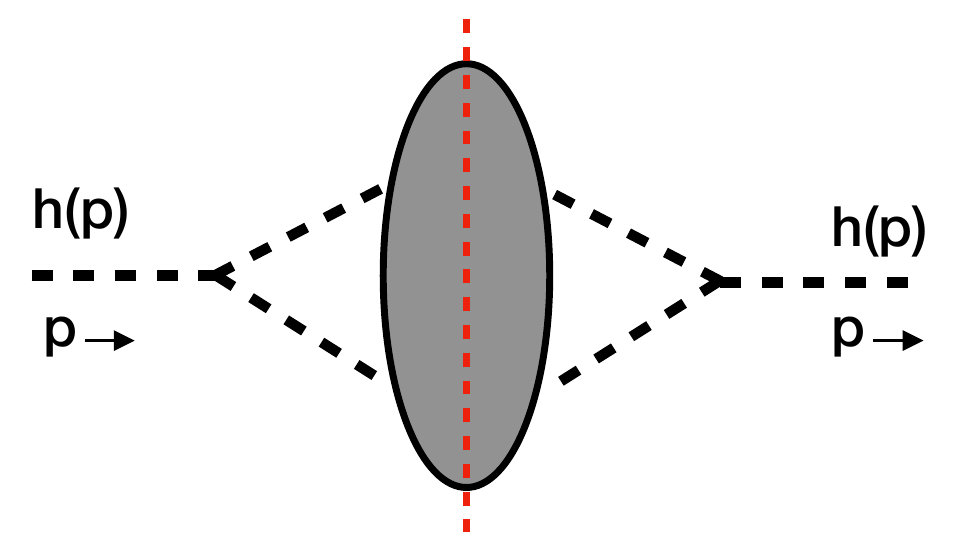}
\hspace{0.5 cm}
\includegraphics[width=0.5\textwidth]{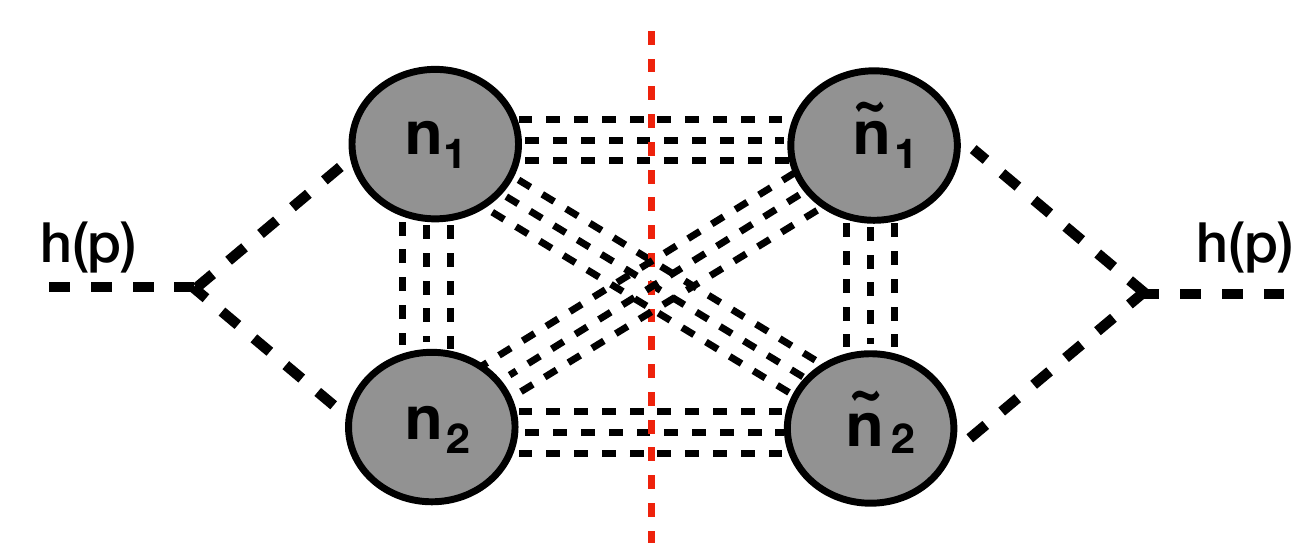}
\end{center}
\caption{Dominant contributions to the self-energy of the Higgs boson from
interference terms between sub-amplitudes for all possible combinations of $n_1, n_2, \tilde{n}_1$ and $\tilde{n}_2$, where $n_1 + n_2 = \tilde{n}_1 + \tilde{n}_2 = n$. Such diagrams contain only multi-particle cuts in the `$t$-channel'.}
\label{fig:V1}
\end{figure}

\begin{figure*}[t!]
\begin{center}
\includegraphics[width=0.35\textwidth]{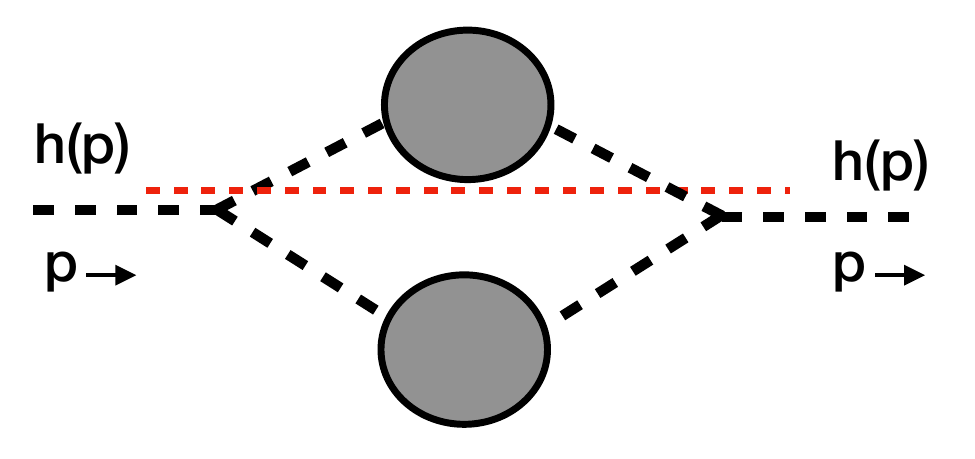}
\hspace{0.5 cm}
\includegraphics[width=0.5\textwidth]{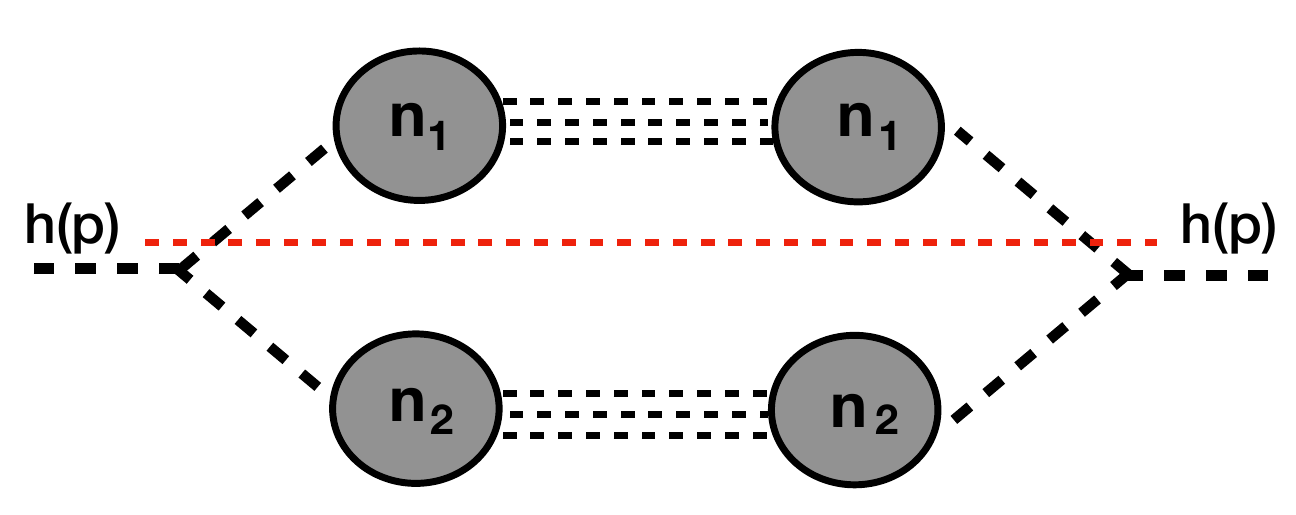}
\end{center}
\caption{Subleading contributions to $\Sigma_h (p^2)$ with no interference between the sub-amplitudes. These non-interference diagrams can be cleanly separated into mutually independent dressed propagators in the loop. These diagrams are 2-particle reducible in the  `$t$-channel'.}
\label{fig:V2}
\end{figure*}

We first take $X$ to be the Higgs field itself and recall the rational for the Higgsplosion of ${\rm Im} \Sigma_h (p^2)$
at $p^2 = E_{*}^2$. The main point we want to emphasise here is that the dominant contribution to the higgsploding
self-energy, or equivalently, the multi-particle decay rate $\Gamma_n$, comes from summing over the interference terms between different diagrammatic contributions to the amplitudes. In particular, each amplitude with $n$ Higgs bosons in the final state
contains of the order of $n!$ terms. The decay rate or the imaginary part of $\Sigma$ arises from squaring the amplitude, then dividing by the symmetry factor of $n!$ and integrating over the phase space. This implies that there are 
$ \sim n! \times n! \times \frac{1}{n!}$ terms. This results  in $\Gamma_n \sim n!$. Clearly, this factorial growth of the rate is 
entirely due to the interference terms (i.e. all the cross terms) in the product of two amplitudes. If, on the other hand, 
one would decide to neglect all the cross terms in the product of two amplitudes, each of which contains 
$n!$ terms, ${\cal A}_n \sim n!$, one would get the total 
of only a single factor of $n!$ which is then cancelled by the $1/n!$ symmetry factor.
In other words, schematically we have,
\[
{\rm Im} \,\Sigma_n \, \sim\, \frac{1}{n!}  \, \left({\cal A}_n \right)^2 
\, \sim\, 
\begin{cases}\,\,
\frac{1}{n!} \times n! \times n! \,\,\sim n! & :\,\,{\rm all\, terms\, included}\\
\,\,\frac{1}{n!} \times n!  \,\,\sim 1 & :\,\,{\rm no\, interference\, terms}\,.
\end{cases}
\label{Sigma_crossterms}
\]
We thus are led to a retrospectively obvious conclusion that Higgsplosion is a result of taking into account all interference 
effects between individual diagrams. These diagrams are sketched in Fig.~\ref{fig:V1} and correspond to the sum of all possible combinations of $n_1,n_2,\tilde{n}_1$ and $\tilde{n}_2$, where $n_1 + n_2 = \tilde{n}_1 + \tilde{n}_2 = n$. This is to be compared with the
diagrams depicted in Fig.~\ref{fig:V2}, where the cross terms between the ${\cal A}_{n_1}$ and ${\cal A}_{n_2}$ sub-amplitudes 
on the left and on the right of the cut were not included. As a result, the diagrams in Fig.~\ref{fig:V2} are subleading relative to
those in Fig.~\ref{fig:V1}, and do not lead to Higgsplosion.

What does lead to Higgsplosion is the correct accounting of  the interference effects in the product of the two amplitudes.
For reader's convenience and for future reference we will now also present a more technical rendering of the above $n!$-counting
argument
for Higgsplosion from intereference, based on the technique of generating $n$-point amplitudes from classical solutions. 
(Readers already familiar with this argument can directly skip to the next section \ref{sec:2.2.2}.)

\bigskip

\centerline{\it A more technical argument: ${\cal A}_{h^*\to n\times h}$ from classical solutions}

\bigskip
\noindent At tree-level, all $n$-point scattering amplitudes for an off-shell field $h^*$ to produce $n$ Higgs particles, 
\[h^* \, \to \, n\times h\,, 
\label{eq:hnh}
\]
can be obtained from solving the Euler-Lagrange equations and
following the generating functions formalism of Brown~\cite{Brown:1992ay}.
For simplicity, as in Ref.~\cite{Khoze:2017tjt}, we will assume a simplified model description of the Higgs boson in terms of a single real scalar field $h$ with the VEV $v$
and the self-coupling $\lambda$,
\[ {\cal L} \,= \, \frac{1}{2}\, \partial^\mu h \, \partial_\mu h\, -\,  \frac{\lambda}{4} \left( h^2 - v^2\right)^2
\,. \label{eq:LSSB} \]
According to \cite{Brown:1992ay}, since
the final state in \eqref{eq:hnh} contains only the outgoing particles, the solution $h_{\rm cl} (x)$ relevant to the problem at hand
should contain 
only the positive frequency modes, $e^{+in M_h t}$ where $M_h = \sqrt{2\lambda}\,v$ is the Higgs boson mass.  
This specifies the initial conditions, or equivalently the analytic structure of
the solution -- its time-dependence is described by the complex variable $z$, 
\[
z(t)\,=\,z_0 \, e^{iM_h t}\,,  \quad M_h\,=\, \sqrt{2\lambda}v\, 
\]
on which the configuration $h_{\rm cl}$
depends holomorphically, so that there is no dependence on the complex-conjugate variable $z^*$,
\[
h_{\rm cl} (\vec{x},t) \,=\, v\,+\, \sum_{n=1}^{\infty} d_n(\vec{x})\, z(t)^n
\, ,
\label{sol-T}
\]
and $d_n(\vec{x})$ are the coefficients of the Taylor expansion in powers of $z$.
Next we consider the simplest kinematical configuration, where all the final state particles are produced at their mass threshold 
(i.e. with vanishing spacial momenta). In this case,
the classical solution in question, $h_{\rm cl}$, is uniform in space and the Euler-Lagrange equation
becomes an ordinary differential equation,
\[
d_t^2 h \,=\, -\lambda\,h^3 +\lambda v^2\,h
\,,
\label{cleq-SSB}
\]
with the initial conditions,  $h_{\rm cl} = v + z + {\cal O}(z^2)$. The coefficients $d_n$ of the Taylor expansion of the classical
solution now become space-independent with $d_1=1$ and the solution is uniquely specified.
Its analytic form is remarkably simple \cite{Brown:1992ay},
\[
h_{\rm cl} (t) \,=\, v\,\frac{1+\frac{z(t)}{2v}}{1-\frac{z(t)}{2v}} \,,
\label{sol-SSB}
\]
and its Taylor expansion reads,
\[
h_{\rm cl} (t) \,=\,  v\,+\, z\, +\, \sum_{n=2 }^{\infty} d_n \, z^n
\,, \quad {\rm with} \quad d_n=(2v)^{1-n}\,, \,\, {\rm for} \,\, n=1,\ldots \infty
\, .
\label{gen-funh}
\]
The presence of the singularity of \eqref{sol-SSB} at $z=2v$
is the consequence of the finite radius of convergence of the Taylor expansion \eqref{sol-SSB}.
The classical solution
$h_{\rm cl}$ defines the generating functional for the tree-level scattering amplitudes.
All $n$-point tree-level amplitudes at threshold are found by differentiating $h_{\rm cl}$ with respect to $z$,
\cite{Brown:1992ay}
\[
{\cal A}_{h^*\to n\times h}\,=\, 
\left.\left(\frac{\partial}{\partial z}\right)^n h_{\rm cl} \,\right|_{z=0}
\,=\, n!\, d_n
\,=\, n!\, (2v)^{1-n}
\,.
\label{eq:amplnh}
\]
The expression \eqref{eq:amplnh} is an exact result and it makes it clear that 
the $1^*\to n$-point amplitudes evaluated on the $n$-particle mass thresholds  grow factorially with 
the number of particles in the final state. The $n!$ behaviour is the consequence of coherently adding 
contributions from the order-$n!$ of Feynman diagrams contributing to these amplitudes.

A remarkable fact that plays an important role in Higgsplosion is that the $n!$ growth of the $n$-point 
amplitudes \eqref{eq:amplnh}
continues to persist in the more general kinematics when the external lines are taken off the mass threshold, and 
furthermore when the leading order quantum corrections from the resummed loops are taken into account.
For the model \eqref{eq:LSSB} the result is
\[
{\cal A}_{h^*\to n\times h} (p_1 \ldots p_n)  \,=\, n!\,  (2v)^{1-n}\,\exp\left[-\frac{7}{6}\, n\varepsilon
\,+\, \frac{\sqrt{3}}{8\pi}\, n\, \lambda n\right]
\,.
\label{eq:ampepsl}
\]
The above expression is derived in the non-relativistic limit, where $\varepsilon$ denotes the kinetic energy per particle per mass
in the final state, $\varepsilon= (E-nM_h)/(nM_h)$, and is taken to be small, $\varepsilon \ll 1$. Hence the first term in the
exponent,
\[ -\, \frac{7}{6}\,  n\,\varepsilon (p_1 \ldots p_n)\,=\, -\, \frac{7}{6}\, \,\frac{1}{2}\,\frac{1}{ M_h^2}\, \sum_{i=1}^{n} \vec{p_i}^2\,,
\]
describes the amplitude dependence on the momenta of non-relativistic particles in the final state; and the second term
is the resummed leading-order loop-level correction. The expression in the exponent (over $n$) 
would also contain higher-order corrections
in $\varepsilon$ and in higher powers of $\lambda n$ which we have neglected.
The details of the derivation of \eqref{eq:ampepsl} and an overview can be found in Ref.~\cite{Khoze:2014kka} and a selection of the
earlier fundamental papers on $n$-point amplitudes in a scalar QFT is 
\cite{Brown:1992ay,Argyres:1992np,Voloshin:1992rr,Voloshin:1992nu,Smith:1992rq,Libanov:1994ug,
Son:1995wz,Voloshin:1994yp,Libanov:1997nt}.

One can now proceed to square the amplitudes \eqref{eq:ampepsl} and integrate them over the $n$-particle phase space 
in the non-relativistic approximation, as was done in e.g. \cite{Son:1995wz,Khoze:2014kka}, ultimately 
providing the foundation of the 
Higgsplosion phenomenon, as explained in Ref.~\cite{Khoze:2017tjt} and analysed further in \cite{Khoze:2017ifq}.
The main point for us here is that Higgsplosion is driven by the total of $n!$ interference terms 
in the product of the amplitudes, as signified by the upper line in \eqref{Sigma_crossterms}.

\bigskip

These considerations imply that the effect of Higgsplosion does not arise from the diagrams with cleanly separated
(i.e. mutually independent) dressed propagators in the loop. These are precisely the diagrams in Fig.~\ref{fig:V2} which
neglect all the interference effects between the upper and the lower half. The Higgslposion effect cannot be derived from 
diagrams containing only the dressed propagators connected by bare vertices, i.e. by working order by order 
 in the loop expanded perturbation theory. What does generate the Higgsplosion 
are the fully interacting diagrams in Fig.~\ref{fig:V1}, with no easily separable dressed propagators and the entire interaction represented by the left diagram in Fig.~\ref{fig:V1}.

\bigskip

\subsubsection{Higgsplosion in the self-energy of other light degrees of freedom}
\label{sec:2.2.2}

\noindent
Let us now consider the self-energy of other Standard Model degrees of freedom. 
More generally, we assume that the field $X$ has a mass much smaller than the Higgsplosion scale,
it can for example be of the order of the electroweak scale, or even lighter, and that it interacts with the Higgs sector.
Does the imaginary part of $\Sigma_X (p^2)$ become large and higgsplodes at some high critical energy scale  $E_{*}$?

\begin{figure*}[t!]
\begin{center}
\includegraphics[width=0.4\textwidth]{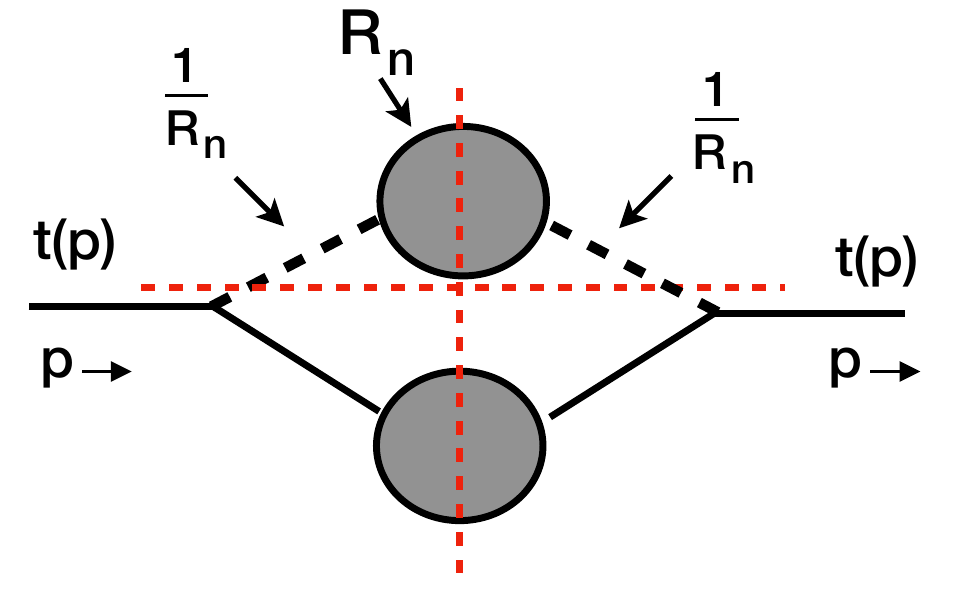}
\hspace{0.5 cm}
\includegraphics[width=0.41\textwidth]{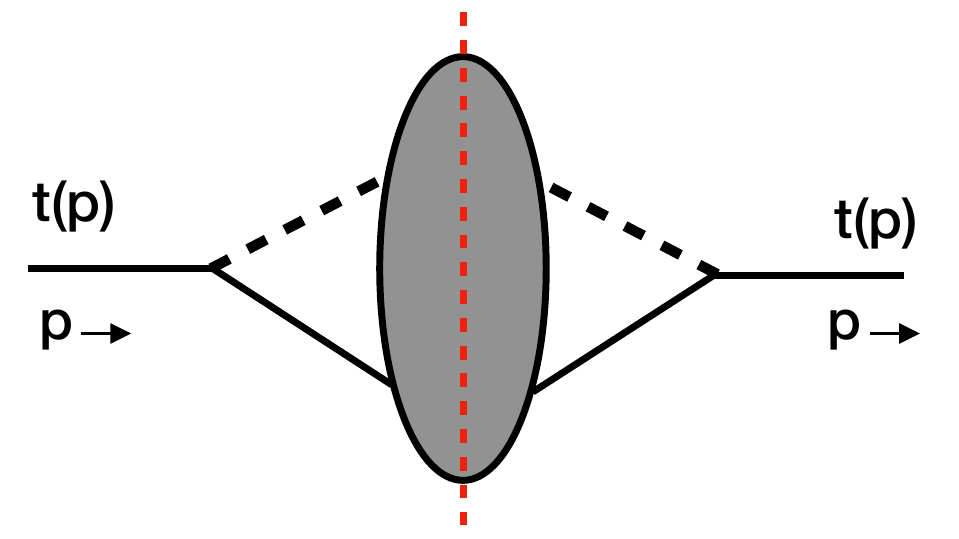}
\end{center}
\caption{Contributions to $\Sigma_t (p^2)$ from mutually independent dressed propagators in the loop.
These sub-processes do not contribute to Higgsplosion and correspond to 2-particle reducible diagrams in the  `$t$-channel'. On the right, dominant contributions to the self-energy of the top quark come from the
interference terms between the sub-amplitudes.
Such diagrams contain only multi-particle cuts in the `$t$-channel'.}
\label{fig:V3}
\end{figure*}

\begin{center}
\begin{figure*}[htp]
\includegraphics[width=0.4\textwidth]{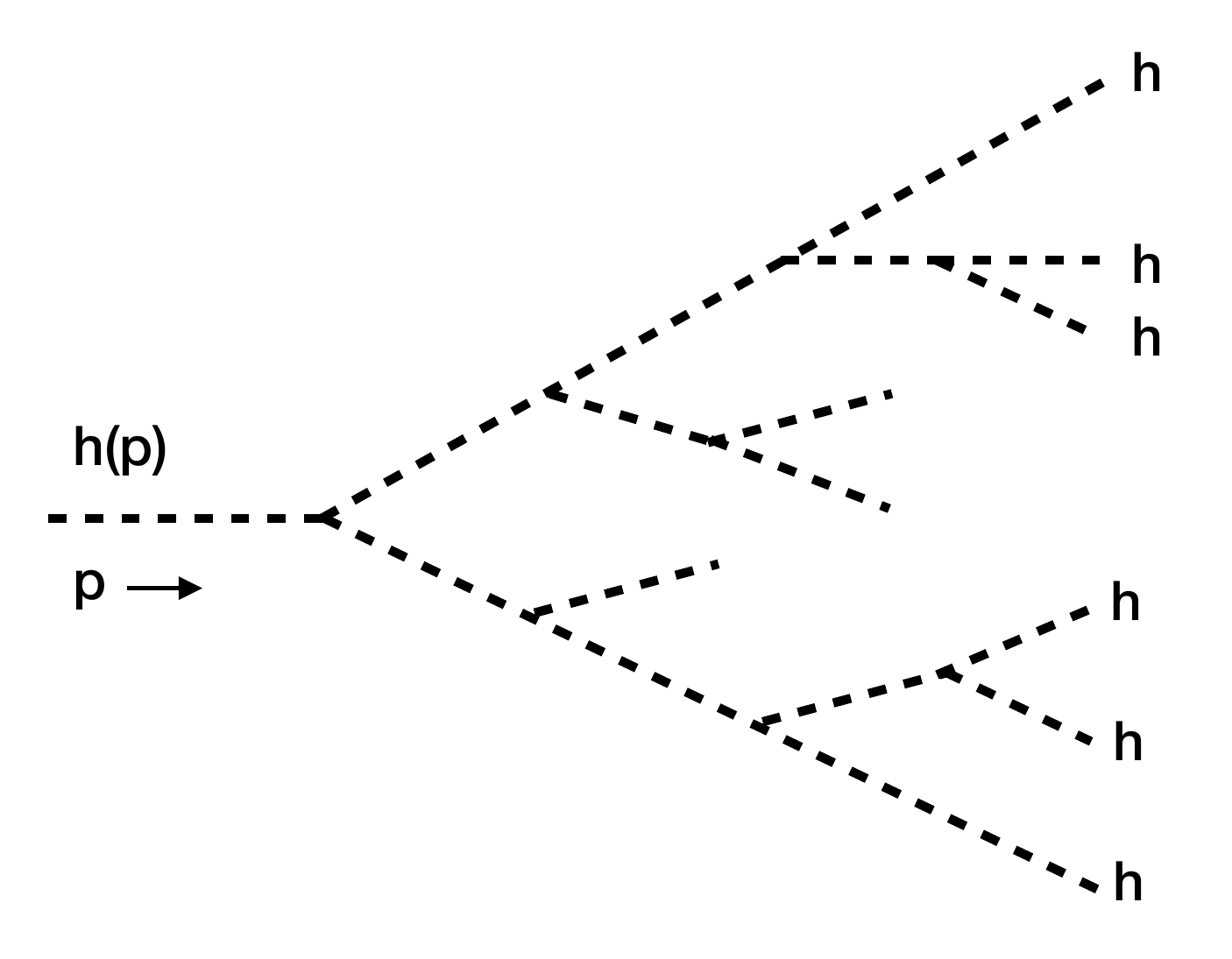}
\hspace{0.5cm}
\includegraphics[width=0.4\textwidth]{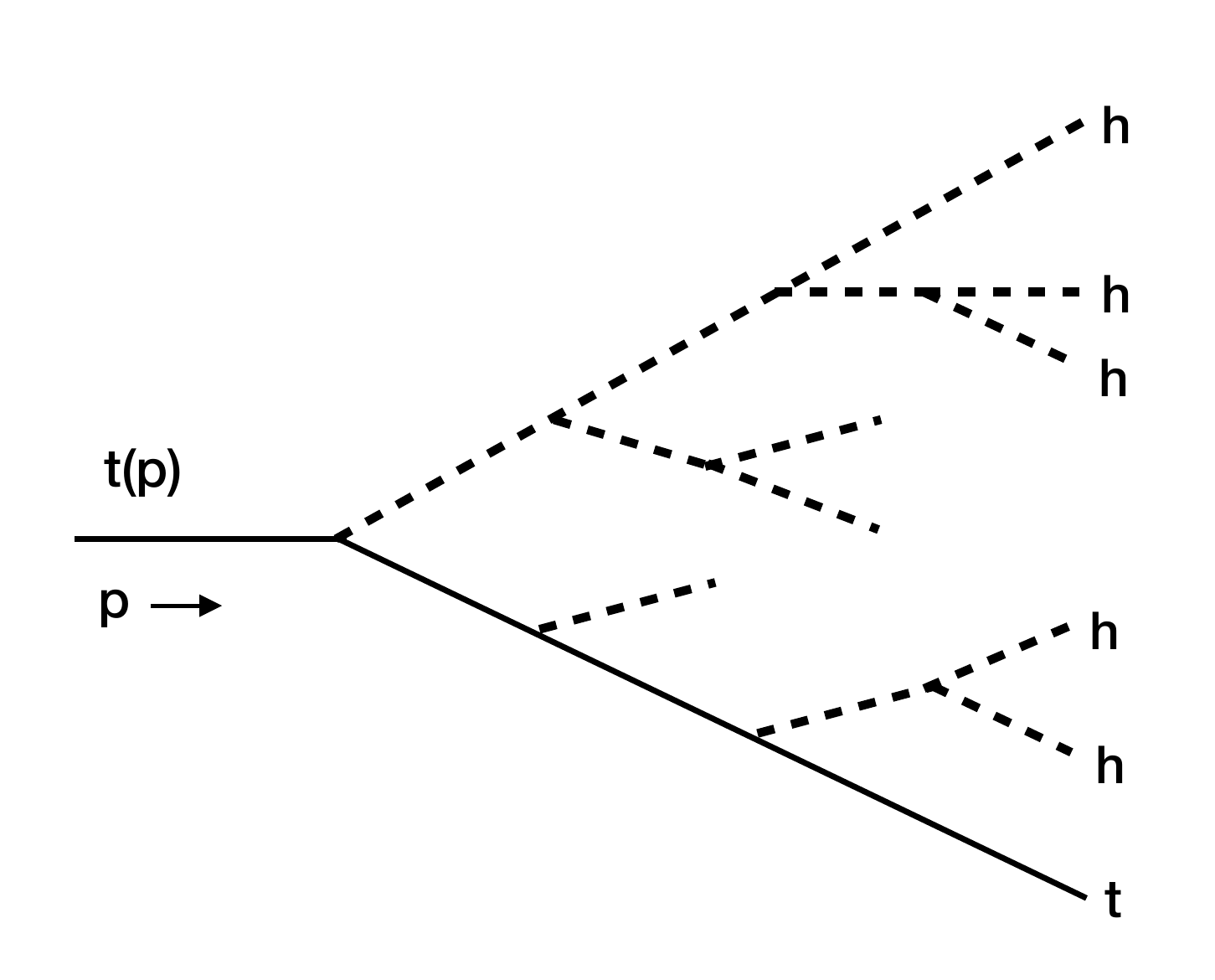}
\caption{Not much difference pictorially between emitting multiple Higgses from the top or from the Higgs internal line}
\label{fig:V5}
\end{figure*}
\end{center}

\noindent For concreteness we first consider here the case of the top quark, $X=t$, but the same qualitative conclusion  
can immediately be drawn for all Standard Model particle (such as the electro-weak vector bosons, gluons and fermions) as
well as other not-too-heavy BSM degrees of freedom coupled to the Higgs.

\bigskip

\centerline{\it Higgsplosion in the self-energy of the top}

\bigskip

For the case of the $t$ quark we concentrate on its self-energy $\Sigma_t(p^2)$ and consider the Yukawa interactions,
$y_t\, \bar{t} t h$ as well as the Higgs self-interactions.
A priory it may be tempting to organise perturbative contributions to $\Sigma_t(p^2)$ in terms of a loop assembled
two or more mutually independent dressed propagators of the Higgs field and of the top quark, connected by bare vertices. 
The leading-order one-loop contribution involving the Higgs and the top dressed propagators is shown on the left in Fig.~\ref{fig:V3}. This amplitude receives a contribution $\sim {{\cal R}_n}$ from Higgsplosion and two contributions $\sim 1/{{\cal R}_n}$ due to Higgspersion and is thus negligible for $p^2 > E^2_*$.

However, following the discussion above, we know that such diagrams ignore the interference terms between 
the sub-amplitudes of the top and the bottom parts of the diagram. To achieve the Higgsplosion effect one should instead consider the more general diagram shown 
in  Fig.~\ref{fig:V3} on the right, which accounts for the contributions of all cross-terms in the product of individual sub-amplitudes; in this sense they are similar to the Higgs self-energy contributions on Fig~\ref{fig:V1}.
Pictorially there is certainly not much difference between the multiple Higgs emissions from the top quark internal
line and the Higgs internal line, as Figure~\ref{fig:V5} indicates. Hence it is entirely likely (and expected) that the imaginary part
of the top quark self energy will also higgsplode in analogy with the pure Higgs case.

\bigskip

To demonstrate the connection between Higgsplosion for the top quark self-energy and the original Higgs field Higgsplosion,  let us compute the tree-level amplitude 
$ {\cal A}_{\,t^* \to \, n\times h + t}$ 
for the top-quark-initiated process depicted in the plot on the right in Fig.~\ref{fig:V5}. We will first compute the generating functional of 
all such amplitudes on the multi-particle mass-threshold by solving the classical equation for the top quark field $\psi_{\rm cl}(t)$ 
\[
\left(i \gamma^0 d_0 \,-\, \frac{m_t}{v}\, h_{\rm cl} \right) \psi_{\rm cl} \,=\,0\,,
\label{eq:tcleq}
\]
in the background Higgs field $h_{\rm cl}(t)$ given by the Brown's solution \eqref{sol-SSB}
that follows from the scalar field Lagrangian, ignoring the back reaction from the $t \bar{t}$ pairs $\psi$ and $\bar{\psi}$ in
\[
{\cal L} \,= \, \frac{1}{2}\, \partial^\mu h \, \partial_\mu h\, -\,  \frac{\lambda}{4} \left( h^2 - v^2\right)^2\,+\,
\bar{\psi}\left(i \gamma^\mu d_\mu \,-\, \frac{m_t}{v}\, h \right) \psi
\,.
\label{eq:LSSBt}
\]
The general procedure is to solve the coupled Euler-Lagrange equations for the Higgs field $h_{\rm cl}(t)$ and 
for the top quark field $\psi_{\rm cl}(t)$, both represented in the form of the double Taylor expansion in terms of the
complex variables $z$ and $\xi^2$,
\begin{eqnarray}
h_{\rm cl} (t) &=&  v\,+\, \sum_{n=0 }^{\infty} \sum_{k=0 }^{\infty}\,d_{n,2k}\,\, z^n \xi^{2k}\,, \quad {\rm with}\quad
d_{0,0}=0\,,\,d_{1,0}=1\,,
\label{gen-funh2}\\
\psi_{\rm cl} (t) &=& \quad \xi \,\,\sum_{n=0 }^{\infty} \sum_{k=0 }^{\infty}\,b_{n,2k}\,\, z^n \xi^{2k}\,, \quad {\rm with}\quad
b_{0,0}=1\,,
\label{gen-funpsi2}
\end{eqnarray}
where
\[ z(t)\,=\,z_0 \, e^{iM_h t}\,,  \quad {\rm and} \quad \xi(t)\,=\,\xi_0 \, e^{im_t t}\,.
\]
$m_t$ is the top mass and the factors of $\xi$ account for the production of top quarks.
For example, in the fermionic generating function \eqref{gen-funpsi2} the factors of $\xi\ \xi^{2k}$ correspond to
a single top plus $k$ additional $t\bar{t}$ pairs in the final state produced from the virtual incoming top quark.
The amplitudes initiated by the virtual Higgs or by the virtual top are obtained by differentiating
respectively $h_{\rm cl}(z,\xi)$ or $\psi_{\rm cl}(z,\xi)$ with respect to the variables $z$ and $\xi$.
The same idea was used previously in the Gauge-Higgs system in Refs.~\cite{Khoze:2014zha,Khoze:2014kka} for computations of amplitudes 
containing Higgses and vector bosons in the final state. Here we are doing the same for the Higgs-$t\bar{t}$ system
\eqref{eq:LSSBt}.

However for our present purpose, which is to account for the amplitudes with a {\it single} top plus multiple Higgses in the final state,
{\it cf.} the plot on the right in Fig.~\ref{fig:V5}, we can neglect the effects of the additional $t\bar{t}$ pair production,
by setting $k=0$ in the sums in \eqref{gen-funh2}-\eqref{gen-funpsi2}.
Hence, as stated earlier, it is sufficient to solve the Dirac equation \eqref{eq:tcleq} in the background of the 
already determined Higgs solution
\eqref{sol-SSB} or \eqref{gen-funh},
\[
h_{\rm cl} \,=\,  v\left( 1\,+\,  2 \sum_{n=1 }^{\infty} (2v)^{-n} \, z^n\right)
\,, \label{gen-funh22}
\]
and search for the top-quark solution in the form,
\[
\psi_{\rm cl}  \,=\, \xi\,\left( 1\,+\,   \sum_{n=1 }^{\infty} \,b_n \, z^n\right)
\,. \label{gen-funh22}
\]
The more general procedure of solving the coupled system in \eqref{gen-funh2}-\eqref{gen-funpsi2}
can also be carried out, as in \cite{Khoze:2014zha}, but we will not pursue it further here. 

We can now solve the classical equation \eqref{eq:tcleq} for the fermionic generating functional $\psi_{\rm cl}$
recursively and determine the Taylor expansion coefficients $b_n$ in \eqref{gen-funh22}. To this end we first act on
the equation \eqref{eq:tcleq} from the left by the operator 
$\left(i \gamma^0 d_0 \,+\, \frac{m_t}{v}\, h_{\rm cl} \right)$ to obtain,
\[
\left(-d_t^2 \,-\, \left(\frac{m_t}{v}\, h_{\rm cl}\right)^2 \,-\, i \gamma^0 \,\frac{m_t}{v} \, (d_t h_{\rm cl})\right) \psi_{\rm cl} \,=\,0\,.
\label{eq:tcleq2}
\]
Finally re-writing the last term in the brackets in the equation above with the help of the original equation
\eqref{eq:tcleq}, we obtain the equation for $\psi_{\rm cl}$ which contains no $\gamma$ matrices
\[
\left(-d_t^2 \,-\, \left(\frac{m_t}{v}\, h_{\rm cl}\right)^2 \,+\,  \frac{(d_t h_{\rm cl})}{h_{\rm cl}}\,d_t\right) \psi_{\rm cl} \,=\,0\,.
\label{eq:tcleq3}
\]
We now define the mass ratio parameter 
\[ \rho\,:=\, m_t/ M_h\,\simeq\, 1.38\,,
\label{eq:rhotdef}
\]
and introduce the rescaled dimensionless variable $t=M_h t$, so that
$z=z_0 e^{it}$ and $\xi = \xi_0 e^{i\rho t}$, and the dimensionless field variable $\phi_{\rm cl}(z)$,
\[
h_{\rm cl} \,=\, v\,(1+2\phi_{\rm cl}) \,, \quad {\rm where}\quad
\phi_{\rm cl}(z) \,=\, \sum_{n=1 }^{\infty}  \, \left(\frac{z}{2v}\right)^n.
 \label{eq:resch2}
 \]
While the Taylor coefficients of the scalar field $\phi_{\rm cl}(z)$ are already fixed  $=1/(2v)^n$ by the known Higgs solution in
\eqref{eq:resch2},
the coefficients $b_n$ defining the fermionic function  $\psi_{\rm cl}(z)$ in \eqref{gen-funh22} are still to be determined
by solving the equation \eqref{eq:tcleq3} 
which takes the form,
\[
\left(-d_t^2 \,-\, \rho^2\, (1+2\phi_{\rm cl} )^2\,+\,  \frac{(2d_t \phi_{\rm cl})}{1+2\phi_{\rm cl}}\,d_t\right) 
\sum_{n=0 }^{\infty} \, b_n \, e^{i(\rho+n)t} \,=\,0\,.
\label{eq:tcleq4}
\]
We solved this equation by iterations using Mathematica, first setting $b_0=1$ and determining all $b_{n>1}$.   
For the first 50 coefficients $b_n:= \tilde{b}_n/(2v)^n$ (using the numerical value of the mass ratio given in \eqref{eq:rhotdef})
we have:
\begin{eqnarray} \tilde{b}_n=&&\{1, 2.75,5.17,8.19,11.8,15.9,20.6,25.7,31.3,37.4,44.,51.,58.5,66.4,74.7,\nonumber \\
&&83.4,92.6,102.,112.,122.,133.,144.,156.,168.,180.,192.,205.,219.,\nonumber \\
&&232.,247.,261.,276.,291.,306.,322.,338.,355.,371.,389.,406.,424.,\nonumber \\
&&442.,460.,479.,498.,518.,538.,558.,578.,599.,620., 641.,\ldots\} \nonumber \\
\label{eq:bns}
\end{eqnarray}
The tree-level scattering amplitudes (or more precisely, the currents) for the process $t^* \to \, n\times h + t$ are then given by
\[
{\cal A}_{\,t^* \to \, n\times h + t}\,=\, 
\frac{\partial}{\partial \xi}\left.\left(\frac{\partial}{\partial z}\right)^n \, \psi_{\rm cl} \,\right|_{z=0}
\,=\, n!\, b_n
\,=\, n!\,  \tilde{b}_n/ (2v)^n
\,,
\label{eq:amplnht}
\]
with the coefficients $\tilde{b}_n$ given in \eqref{eq:bns}. Restoring the kinematic dependence due to $n$ non-relativists 
momenta of the Higgs bosons in the final state, as in  \eqref{eq:ampepsl}, we have 
\[
{\cal A}_{\,t^* \to \, n\times h + t} \, (p_1,\ldots,p_{n+1})\,=\, n!\,\,  \tilde{b}_n \, (2v)^{-n}\, e^{-(7/6) n \varepsilon} \,
\,.
\label{eq:amplnht2}
\]
This amplitude retains the factorial growth with the number of the Higgs bosons in the final state and in view of the
coefficient values in \eqref{eq:ampepsl} with $\tilde{b}_n >1$, the amplitude initiated by the top-quark line
is not inferior to the $n$-point amplitude for the pure Higgs production process
in  \eqref{eq:ampepsl},
\[
{\cal A}_{h^* \to \, n\times h} \, (p_1,\ldots,p_{n})\,=\, n!\,\,  (2v)^{1-n}\, e^{-(7/6) n \varepsilon} 
\,.
\label{eq:amplnh2}
\]
Based on these tree-level considerations, and admittedly not having attempted to add and re-sum higher order quantum corrections
involving top quark loops, we conclude that the Higgsplosion of the top quark self-energy ${\rm Im} \Sigma_t(p^2)$ is as likely as the
Higgsplosion of the Higgs boson  ${\rm Im} \Sigma_h(p^2)$ at $p^2 = E^2_*$.

\bigskip

\centerline{\it Higgsplosion in the self-energy of vector bosons}

\bigskip

As another example of higgsploding the SM degrees of freedom, one can consider the amplitudes involving
the Higgs as well as the weak-sector massive vector bosons. Tree-level amplitudes for multiple Higgs bosons and longitudinal components of $W$'s and $Z$'s were already considered in Refs.~\cite{Khoze:2014zha,Khoze:2014kka}
on and off the multi-particle thresholds. The formalism is very similar and involves solving the time-dependent
classical equations for the Gauge-Higgs system,
\begin{eqnarray}
&&- \, d_t^2 h \,\, =\,   \lambda\,h^3  -\lambda v^2\,h +\frac{g^2}{4} (A_L^a)^2 h
\,,\label{cleq-h3}\\
&&-\,  d_t^2 A_L^a \,=\, \frac{g^2}{4} h^2 A_L^a \,.
\label{cleq-A3}
\end{eqnarray}
The classical solutions for the Higgs field, $h_{\rm cl}$, and for the longitudinal components of the vector boson
fields ${A_L^a}_{\rm cl}$ are represented as double Taylor expansions in terms of the $z$ and the $w^a$ variables,
\[
z(t) \,=\, z_0\, e^{i M_h t}\,,\quad {\rm and}\quad
w^a(t)\,=\, w_0^a\, e^{i M_V t}\,,
\label{eq:zw}
\]
where $M_V$ is the vector bosons mass.
\begin{eqnarray}
h_{\rm cl}(z,w^a) &=& v\,+\, 2v\, \sum_{n=0}^{\infty}\sum_{k=0}^{\infty} \, d_{n,2k}\,\left(\frac{z}{2v}\right)^n\,\left(\frac{w^a w^a}{(2v)^2}\right)^k
\,,\label{eq:dTh-fin}
\\
{A_{L}^a}_{\rm cl}(z,w^a) &=& w^a\,\sum_{n=0}^{\infty}\sum_{k=0}^{\infty} \, a_{n,2k}\, \left(\frac{z}{2v}\right)^n\,\left(\frac{w^a w^a}{(2v)^2}\right)^k
\,,
\label{eq:dTA-fin}
\end{eqnarray}
with the lowest-order Taylor coefficients $d_{0,0}=0$ and $a_{0,0}=1$.

The amplitudes involving vector bosons and Higgs bosons in the final state 
on the multi-particle thresholds are given by the following expressions
in terms of the Taylor expansion coefficients $d_{n,2k}$ and $a_{n,2k}$,
\[
  {\cal A}_{h^* \to n\times h + m\times Z_L}\,= \, (2v)^{1-n-m}\, n!\, m!\, \,d_{n,m} \,,
  \label{Ah-fin}
 \] 
 and for the longitudinal Z decaying into $n$ Higgses and $m+1$ vector bosons we have,
 \[
 {\cal A}_{Z_L^* \to n\times h + (m+1)\times Z_L}\,= \, \frac{1}{(2v)^{n+m}}\, n!\, (m+1)!\, \, a_{n,m}\,,
  \label{AZ-fin}
 \]
 and obtained by differentiating the classical generating functions in \eqref{eq:dTh-fin} and
 \eqref{eq:dTA-fin}
 with respect to the variables $z$ and $w^a$ whre $a=1,2,3$ is the isospin index.

The coefficients $d_{n,m}$ and $a_{n,m}$ were computed in Ref.~\cite{Khoze:2014zha} by solving the 
classical equations above for the given ratio of the vector to the Higgs boson masses,
\[
\kappa :=\, \frac{g}{2\sqrt{2\lambda}} \,=\,  \frac{M_V}{M_h} \,\simeq\, 0.64\,.
\label{eq:kappadef}
 \]
In particular, for the simplest case of $m=0$ of no vector boson pairs present in the final state,
the amplitudes for the  $Z_L^* \to n\times h +  Z_L$ process are given by
 \[
 {\cal A}_{Z_L^* \to n\times h + \times Z_L}\,= \, \frac{1}{(2v)^{n}}\, n!\, \, a_{n}\, e^{-(7/6) n}  \,,
  \label{AZ-fin}
 \]
 with the first 50+ coefficients $a_n = a_{n,0}$ given by
\begin{eqnarray} 
a_n=&&\{1,0.718,0.678,0.652,0.633,0.617,0.605,0.594,0.585,0.577,0.57,0.564,0.558,\nonumber \\
&&0.553,0.548,0.543,0.539,0.535,0.531,0.528,0.524,0.521,0.518,0.516,0.513,0.51,\nonumber \\
&&0.508,0.505,0.503,0.501,0.499,0.497,0.495,0.493,0.491,0.489,0.488,0.486,\nonumber \\
&&0.484,0.483,0.48,0.479,0.477,0.476,0.474,0.481,0.48,0.479,0.477,0.476,\nonumber \\
&&0.474,0.473, 0.472,0.471,0.469,0.468,0.467,0.466,0.465,0.464,0.463, \ldots\} \nonumber \\
\label{eq:an50}
\end{eqnarray} 
Given the $\sim 1$ values of the coefficients in \eqref{eq:an50}, the amplitudes \eqref{AZ-fin}
with a single vector boson line are numerically very similar to the pure Higss amplitudes 
in \eqref{eq:amplnh2}. These considerations imply that the Higgsplosion should also occur in the self-energy of the 
vector boson propagator with the Higgsplosion scale of the same (or similar) magnitude as $E_*$ in the pure Higgs case.

\bigskip

While the calculation of  higher order quantum corrections to the tree-level amplitudes involving interactions with the top quark and the vector 
bosons  have not been carried out so far, which could certainly change the scale where Higgsplosion occurs quantitatively, it is nevertheless a self-consistent 
assumption to conjecture that the self-energy will higgsplode for all light degrees of freedom coupled to the Higgs boson.

\bigskip

\centerline{\it Higgsplosion in the self-energy of the graviton}

\bigskip

We start with the Lagrangian 
\[ 
\sqrt{-g}\, {\cal L} \,=\, \sqrt{-g} \left( -\frac{M_{\rm Pl}^2}{2} \, R\,+\, {\cal L}_{\rm matter}\,+\, {\cal L}_{\rm GF} \right)\,,
\]
where he first term is the Einstein-Hilbert Lagrangian of gravity, the second term is the coupling of gravity
to the Higgs field $h$, and the final term is the gauge-fixing term for gravity.
$R$ is the scalar curvature and
$M_{\rm Pl}$ is the reduced Planck mass, which is often re-written as $M_{\rm Pl}=2/\kappa$, 
where $\kappa^2=32 \pi G$ and $G$ is the
Newton's constant $G=  (1.22 \times 10^{19}\, {\rm GeV})^{-1}$.
The graviton field ${\tt g}_{\mu\nu}$ is defined as the fluctuation of the metric tensor around the Minkowski space
metric,
$g_{\mu\nu}(x)\,=\, \eta_{\mu\nu} \,+\, \kappa\, \chi_{\mu\nu}(x)$ and we use the convention $ \eta_{\mu\nu}=(+1,-1,-1,-1)$.
We can now simplify this Lagrangian by retaining only the terms linear and quadratic in the graviton field
$\chi_{\mu\nu}$, which takes the form ({\it cf.} e.g. the lecture notes \cite{Donoghue:2017pgk}),
\[ 
{\cal L} \,=\, \frac{1}{2} \,\chi_{\mu\nu}\, P^{\, \mu\nu\alpha\beta}\, \partial^2 \, \chi_{\alpha\beta}\, -\, 
\frac{\kappa}{2} \, \chi^{\mu\nu}\, T_{\mu\nu}
\,.
\]
Here 
\[ P^{\, \mu\nu\alpha\beta}\,=\, \frac{1}{2} \left( \eta^{\mu\alpha}\eta^{\nu \beta}
+\eta^{\mu\beta}\eta^{\nu\alpha}-\eta^{\mu\nu}\eta^{\alpha \beta}\right)\,,
\]
and $T_{\mu\nu}$ is the energy-momentum tensor of the Higgs field,
\[
T_{\mu\nu}\,=\, \partial_\mu h\, \partial_\nu h\,-\, \eta_{\mu\nu}  \,{\cal L}(h)\,\]
where ${\cal L}(h)$ is our scalar-field Lagrangian \eqref{eq:LSSB}.
To describe the tree-level process where the incoming graviton decays into a multi-particle final state
made entirely out of Higgs bosons (and no additional gravitons), $\chi_{\mu\nu} \to n \times h$,
it is sufficient to solve the linearised equation for the graviton field in the background of the classical Higgs solution.
The graviton equation reads \cite{Donoghue:2017pgk},
\[
\partial^2 \, P^{\, \mu\nu\alpha\beta}\, \chi_{\alpha\beta}\,=\, \frac{\kappa}{2}\, T^{\mu\nu}(h_{\rm cl})\,.
\]
Furthermore, when we restrict the particles in the final state to be on the multi-particle thrsehold,
the equation becomes an ordinary differential equation with respect to time, $\partial^2  \to\, d_t^2$,
and for the background Higgs field we can as before use the already known analytic expression \eqref{sol-SSB}.

There is an additional simplification, namely it is easy to check that when evaluated on the classical Higgs configuration \eqref{sol-SSB},
the energy-momentum tensor becomes,
\begin{eqnarray}
T_{00}&=& {\cal H}(h_{\rm cl}) \,=\, \frac{1}{2}\, (d_t h_{\rm cl} )^2\, +\,  \frac{\lambda}{4} \left( h_{\rm cl}^2 - v^2\right)^2  \,\,=\,  0\,,
\\
T_{ij}&=& \delta^{i}_{j}\,  {\cal L} (h_{\rm cl}) \,=\, -\, \delta^{i}_{j}\, \frac{\lambda}{2} \left( h_{\rm cl}^2 - v^2\right)^2\,.
\end{eqnarray}
Hence we search for the solution of
\[
d_t^2 \, \tilde\chi_{\rm cl} \,=\, -\,\frac{\kappa\lambda}{4}\, \left( h_{\rm cl}^2 - v^2\right)^2\,
\label{eq:grav}
\]
where $\tilde\chi_{\rm cl}\,:=\, (P\chi_{\rm cl})^{ii}(t)$ with no sum over $i$. 
We are looking for the solution of the form of the Taylor expansion in the variable $z=z_0 e^{i M_h t}$,
\[
\tilde\chi_{\rm cl}(z)\, =\, \frac{v^2}{M_{\rm Pl}}\, \sum_{n=2}^{\infty} \, g_n\, \left(\frac{z}{v}\right)^n\,,
\]
and the Taylor coefficients $g_n$ can be determined analytically by solving \eqref{eq:grav}
by iterations. Working in units of $v=1$ we find,
\[
g_n\,=\, \frac{1}{n^2} \,\left( h_{\rm cl}^2-1\right)^2|_{z^n}\,\,=\, \frac{8}{3}\, \frac{n^2-1}{2^n\, n}\,.
\]
In summary, we have for the generating function of all graviton decay amplitudes into $n$ Higgs bosons 
the following expression,
\[
\tilde\chi_{\rm cl}(z)\, =\, \frac{8}{3}\,\frac{v^2}{M_{\rm Pl}}\, \sum_{n=2}^{\infty} \, \frac{n^2-1}{n} \left(\frac{z}{2v}\right)^n\,.
\]
This is to be compared with the pure Higgs solution in \eqref{gen-funh}. 
In terms of the $h_{\rm cl}$ Taylor coefficients $d_n = (2v)^{1-n}$ we now have
the graviton Taylor coefficients (including now all dimensionfull factors),
\[
g_n \,=\, \frac{v}{M_{\rm Pl}} \, \frac{4}{3} \, \frac{n^2-1}{n} \, d_n\,.
\]
As the $n$-point amplitude is given by 
\[
{\cal A}_{{\rm graviton} \to n\times h} \,=\, n! \, g_n
\]
we see that the graviton decay rate ${\cal R}_n$
will be suppressed by a relative constant factor of $\frac{v}{M_{\rm Pl}}$ times $n^2$.
This suppression is by a constant, i.e. energy and $n$-independent factor, so it will not prevent
Higgsplosion, but will result in the considerably lowered Higgsplosion scale $E_*$ for the graviton relative to Higgses.

\subsubsection{Heavy degrees of freedom}
\label{sec:2.2.3}

The behaviour of the self-energy of the degrees of freedom with masses $M_X \gg E_{*}$ will depend on whether the 
field under consideration is stable or not, in other words whether the heavy $X$ can decay into light degrees for freedom 
$X \to {\rm Light} + n\, h$, or if it is required that $X \to X + n\, h$. One can consider the following possibilities:

\begin{enumerate}
\item{} For a heavy scalar field $X$ which can decay into multiple Higgs bosons alone, $X \to n\,h$, the discussion follows Sec.~5 of  \cite{Khoze:2017tjt}. 
The main point is that there is no difference between $X$ or $h$ on the external lines of the self-energy.
The result is that ${\rm Im} \, \Sigma_X (p^2)$ higgsplodes.\footnote{In fact, the exponential factor in \eqref{eq:Rn} 
is the same for the incoming $X$ 
in the process $X \to n\,h$ and for the virtual Higgs $h^*$ in the process $h^* \to n\,h$.}

\item{} For a bosonic or fermionic $X$ which decays as $X \to {\rm Light} + n\, h$ the situation is very similar to the top quark 
self-energy. Here $X$ can for example be a very heavy $10^{11-13}$ GeV sterile neutrino which first decays into a 
light neutrino and the Higgs, and subsequently into many Higgses. The self-energy of such $X$ is also expected to higgsplode
in direct analogy to the top quark (or other SM fermions).

\item{} For stable heavy degrees of freedom $X$ the story is different because the remaining $X$ in the decay process
$X \to X+n\,h$
will carry away the momentum
of the order of the mass of $X$, thus depleting the energy left for the Higgsplosion into multiple Higgs bosons. Thus Higgsposion could occur 
only at energies $E_*^X \simeq M_X + E_*^{nh} $. If $M_X \gg E_*^{nh}$ such states will reintroduce a Hierarchy problem for the Higgs boson, and hence 
should be avoided in model building.
\end{enumerate}

\medskip

\subsection{Solving the Hierarchy problem}
\label{sec:2.3}

Here we are interested in assessing the contributions to the Higgs self-energy $\Sigma_h (p^2)$
at {\it low} energy scales, of the order of the measured  Higgs mass and much below the Higgsplosion scale, $p^2 \sim M_h^2 \ll E_{*}$.
Clearly the multi-particle Higgs production relevant for Higgsplosion is impossible at such low scales 
and hence the imaginary part of $\Sigma_h(p^2)$ plays no role, we are interested predominantly in how big or small
its real part is after the integration over the loop momenta.

The whole point and the origin of the Hierarchy problem for the Higgs mass for such low values of the external momenta
$p^2 \sim M_h^2$ of the ${\rm Re} \Sigma_h(p^2)$ is that the super-heavy degrees of freedom propagating in the loops 
contributing to $ \Sigma_h(p^2)$,
do not decouple because of the UV divergencies in the integration over the loop momenta. Thus to address the Hierarchy problem
it is sufficient to concentrate only on the UV-divergent and/or $M_X$-sensitive contributions to the self-energy of the Higgs.

The UV-sensitive diagrams contributing to the self-energy at small external momenta are conceptually different from the 
higgsploding contributions which were UV-finite at tree level and computed at $p^2 = E_*^2 \gg M_h$.
In the case at hand one needs to address the cases with the smallest numbers of propagators (and hence the vertices) in the
loops -- after all we are only after the UV-divergent contributions.
In this case the correct procedure is the expansion of the self-energy in the number of loops, each with the minimal number of internal
propagators. The new input from the Higgsplosion is that the internal propagators 
$\Delta_X(p)$ are the dressed propagators \eqref{eq:pfin2F}. It then immediately 
follows that all the UV divergencies and even the finite terms are
cut-off by the values of the loop momenta approaching the Higgsplosion scales of the relevant self-energy factors in the dressed propagators.
For example, following \cite{Khoze:2017tjt} for the case of the heavy scalar $X$  interacting with the Higgs sector via
\[
 {\cal L}_X \,=\, 
 \frac{1}{2} \partial^\mu X \,\partial_\mu X\,-\, \frac{1}{2} M_X^2 \, X^2 \,-\, 
 \frac{\lambda_{P}}{4}\, X^2 h^2 \,-\, {\mu} X h^2\,,
\label{eq:toym2}
 \]
 the 1-loop radiative correction to the Higgs mass parameter is UV-finite and $\ll M_X^2$,
 \[
 \Delta M_h^2 \,\sim\, \lambda_P \int \frac{d^4p}{16\pi^4} \, \frac{1}{M_X^2\,-\, p^2\, +i\, {\rm Im}\, \Sigma_X(p^2)}
\,\propto\, \lambda_P \,\, \frac{E^2_\star}{M_X^2}\,\, E^2_\star \quad \ll \, \lambda_P M_X^2\,.
 \label{eq:Xhp} 
\]

\noindent In general the computation of the real part of $\Sigma_h(p^2)$ at $p^2\simeq M_h^2$ proceeds as
explained in Sec. 5 of our original Higgsplosion paper \cite{Khoze:2017tjt}. 
We note that the diagrammatic technique employed now involves dressed propagators and bare vertices and is conceptually 
different to the 
diagrams in Fig.~\ref{fig:V2} that contributed to the Higgsplosion of the self-energy discussed in the preceding sections.
However, there is no contradiction. In addressing the Hierarchy we are working in a different regime,
where the external momenta are much smaller than the Higgsplosion scale and we are
tracing what used to be the UV-divergent contributions that arose from integrations over the loop momenta. 
Such diagrams are correctly accounted by the
loop diagrams with the minimal numbers of dressed propagators in the loops, and in presence of Higgsplosion,
the loop momenta are dynamically cut off at $E_{*}$. 
Hence there are no contributions to $M_h^2$ proportional to neither the $\Lambda_{UV}^2$ nor
the $\sim M_X^2$ factors. The radiative corrections to the Higgs mass squared are cut off at the much lower scales set by
$E_{*}$, as indicated in \eqref{eq:Xhp}, thus solving the Hierarchy problem.


\bigskip
\subsection{UV Finiteness of n-point functions and Asymptotic Safety}
\label{sec:2.4}
\bigskip

As we already noted in Sec.~\ref{sec:2.1} Higgsplosion is triggered in a given QFT if and when the multi-particle decay widths $\Gamma_n$
of all degrees of freedom become exponentially large above a certain dynamically generated non-perturbative scale $E_*$, and exhibit the behaviour described by \eqref{Sigma_phi}.

\begin{figure}[htp]
\begin{center}
\includegraphics[width=0.35\textwidth]{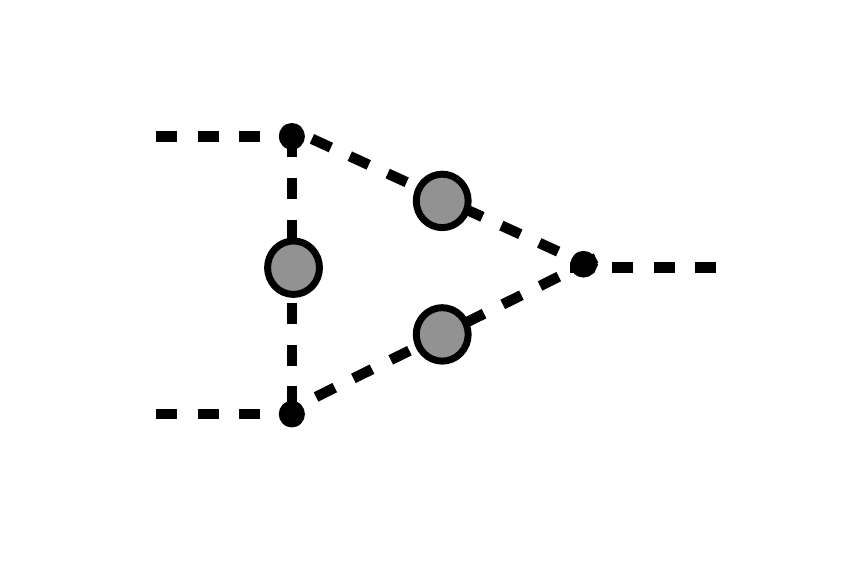}
\hspace{0.5cm}
\includegraphics[width=0.3\textwidth]{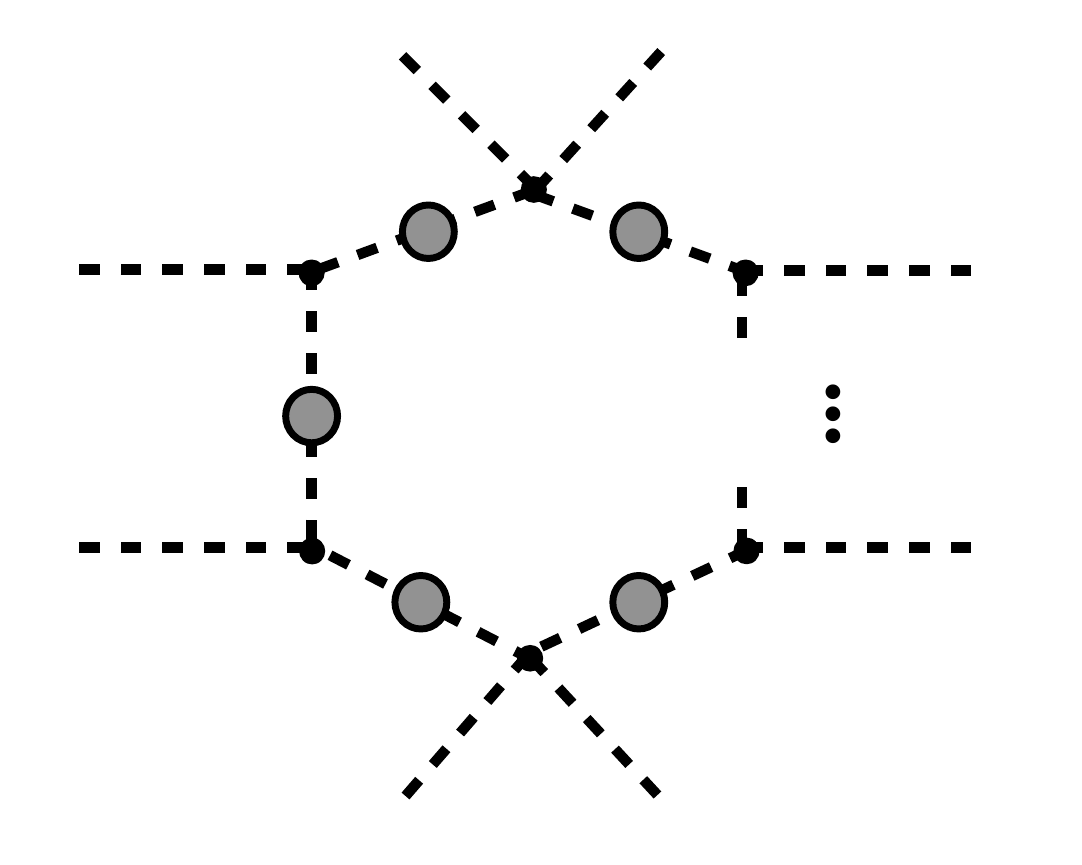}
\caption{One loop contributions to the three-point (left) and $n$-point (right) Green functions. 
The grey blobs represent dressed propagators and the black dots are the microscopic 3- and 4-point  interaction vertices.}
\label{fig:npoint}
\end{center}
\end{figure}

While a fully non-perturbative self-consistent formalism is currently lacking, in this paper we would like to advocate a simple diagrammatic 
approach for computing quantum effects in a Higgsploding QFT based on a resummed perturbation theory. The two building blocks are
(I) the dressed propagators \eqref{eq:pfin2F} that include the Higgspersion formfactor for all field theoretic degrees of freedom 
present in the problem,
and (II) the bare vertices that are read directly from the microscopic Lagrangian.
The renormalised vertices which depend on the RG scale $\mu$ are then obtained in the standard way from computing the $n$-point 
one-particle irreducible LSZ-amputated Green functions $G_n$. These computations are performed order  by order in the loop expansion,
with the only difference from the usual approach that one is required to use the dressed propagators \eqref{eq:pfin2F} on all internal lines. The leading order one-loop contributions to the 3-point and the $n$-point vertices are shown in Fig.~\ref{fig:npoint}.

This way of computing quantum effects in a Higgsploding QFT leads to a powerful conclusion 
that all momenta of virtual particles propagating in the loops are effectively cut off at the Higgsplosion scale $E_*$. 
Integrations over the loop momenta are convergent,
all the contributions to the $n$-point functions are UV finite and
quantum fluctuations are damped above $E_*$.

There is an interesting parallel between this approach and Polchinski's implementation of the Wilson approach to renormalization  
\cite{Wilson:1971bg,Wilson:1973jj} presented in Ref.~\cite{Polchinski:1983gv} for a massive $\phi^4$ theory. 
In the construction of \cite{Polchinski:1983gv} the UV cut-off is implemented
by multiplying the propagators by a formfactor $K(p^2/\Lambda_0^2)$ which is equal to 1 for momenta $p^2\le \Lambda_0^2$
and rapidly vanishes for $p^2 > \Lambda_0^2$. What defines the theory with the (large) UV cut off $\Lambda_0$ is the Lagrangian with the modified propagator and bare vertices. 
When the cut off is
lowered from $\Lambda_0$ to $\Lambda_{\rm R}$, one is required to integrate out the high momentum components of the field.
This is implemented by changing the formfactor in the propagator to $K(p^2/\Lambda_{\rm R}^2)$ and integrating
out the modes with $p^2 >\Lambda_{\rm R}^2$. This generates new effective interactions and expresses them in terms 
of the couplings at the scale $\Lambda_{\rm R}$. 
The analogy of our method for computing the $n$-point functions with the approach of~\cite{Polchinski:1983gv} is that 
the theory with a large UV cut off is defined by the modified propagators and bare vertices. The momentum modes above the
cut-off are switched off in both cases simply by the fact that the modified propagators vanish. In the case of Higgsplosion,
what we referred to as the large UV cut off is the dynamically generated Higgsplosion scale $E_*$, and the original propagators are modified
 by the self-energy $\Sigma (p^2)$ contributions \eqref{eq:pfin2}. The theory with momenta above the Higgsplosion scale 
is the theory above the UV cut-off; its propagators vanish so it has has no propagating degrees of freedom left, but its vertices are 
the usual bare vertices fixed at the scale $E_*$. There are no quantum fluctuations and no running above the scale $E_*$.

The coupling constants in a Higgsploding theory receive no quantum corrections from the modes above $E_*$.
Hence the running couplings become flat for the values of the RG scale $\mu > E_*$. Below the Higgsplosion
scale, $\mu < E_{*}$, the couplings exhibit the 
usual logarithmic running with $\mu$, but above $E_{*}$, their beta functions vanish and the couplings stay constant 
with the values determined at $E_{*}$. This amounts to the asymptotically safe theory.

\bigskip

\subsection{High-energy scattering with Higgspersion}
\label{sec:2.5}
\bigskip

Before concluding this section, let us pose the following question:  is it true that due to Higgsplosion and Higgspersion, all scattering processes beyond the Higgsplosion scale become non-interacting? The couplings stay constant, while all scatterings cross-sections nevertheless vanish exponentially
as dictated by \eqref{eq:poly_ev_odd}. Is this a consistent picture?

\begin{figure}[htp]
\begin{center}
\includegraphics[width=0.7\textwidth]{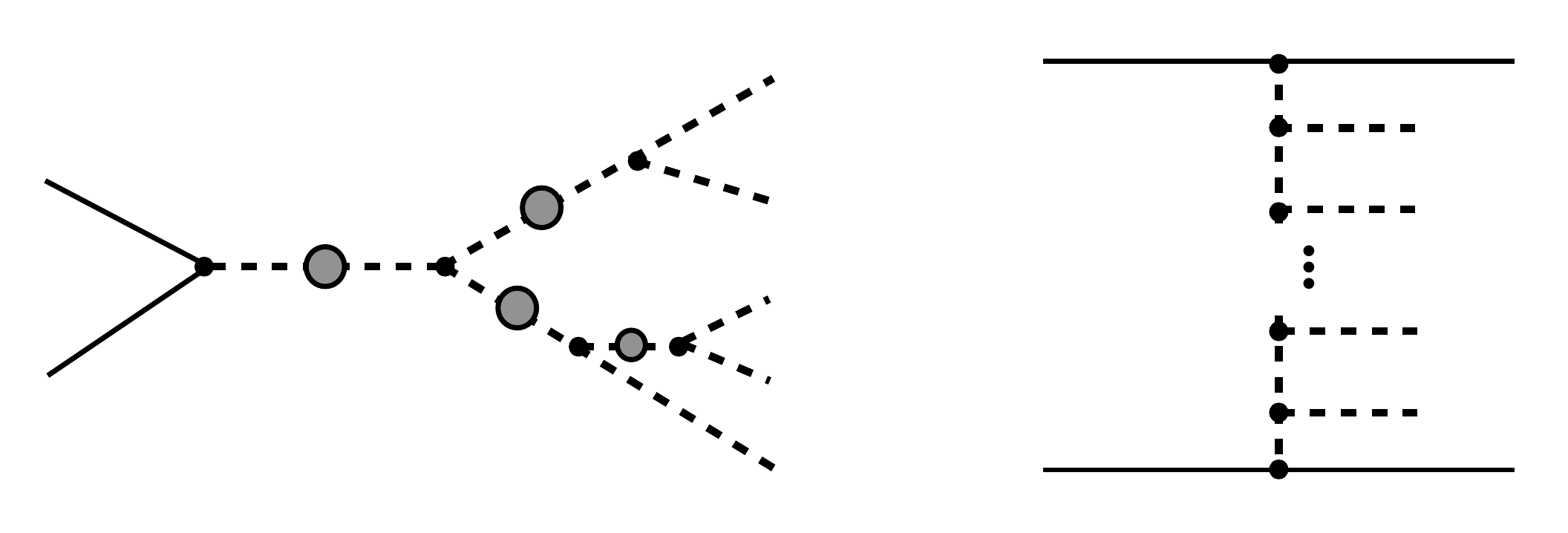}
\caption{$s$-channel (left) and $t$-channel (right) interaction diagrams.}
\label{fig:st}
\end{center}
\end{figure}

The main point is whether or not the internal propagators appearing in the diagrams contributing to the high-energy scattering process
involve momenta that can exceed the Higgsplosion scale $E_*$. For the high-energy processes dominated by the $s$-channel exchanges,
such as the diagram on the left of Fig.~\ref{fig:st}, the presence of at least one propagator with $p^2 \ge E_*^2$ is unavoidable 
when the total $\sqrt{s}\ge E_*$. These are the processess we are considering in this paper and they indeed shut down beyond the
Higgsplosion scale due to the Higgspersion of the dressed $s$-channel propagator, where $\Delta(p^2) \sim 1/{\cal R}(p^2) \to 0$. A physically intuitive meaning of this effect is that the propagating degree of freedom simply disappears from the spectrum, it is no longer an individual particle but a manifestation of multi-Higgs radiation.

On the other hand, not all of the contributing diagrams
to high energy $\sqrt{s}\ge E_*$ scattering are of this form. The diagram on the right of Fig.~\ref{fig:st} represents a $t$-channel process.
The transverse momentum is not required to be large or anywhere near the scale $E_*$. Hence the Higgsplosion/Higgspersion effects 
will be absent in such processes, and their contributions to the scattering processes will survive even at high energies.

In summary, Higgspersion shuts down the processes which probe shorter and shorter distances at higher and higher energy. This is not happening because the couplings 
vanish but because the internal propagators with momenta above $E_*$ in fact turn off and disappear.
The dynamically generated Higgsplosion scale sets the minimal length scale $r_* \sim 1/E_*$ that can be probed at any arbitrary high energy\footnote{In the analysis of Higgsplosion
for the $gg \to h^* \to n\times h$ processes, it is the Higgspersion effect in the first Higgs propagator of the most energetic state $h^*$ that turns off
the cross-section and restores unitarity. The effects of the Dyson resummation of subsequent intermediate Higgs propagators 
are irrelevant at energies just above the Higgsplosion threshold. This is because after each 2-point or 3-point splitting of the virtual
Higgs state $h^*$, the energies/virtualities carried by the emerging propagators are 1/2 or 1/3 of the initial energy; they fall below the 
Higgsplosion scale, and their self-energy insertions are irrelevant.}.

In this sense Higgsplosion provides a dynamical realisation of the idea of classicalization \cite{Dvali:2010jz,Dvali:2016ovn}, 
where the role of the classicalization radius
is played by the Higgsplosion scale $r_*$. 
It also resounds the importance of multi-regge kinematics for high-energy scatterings as introduced by Balitsky, Fadin, Kuraev and Lipatov \cite{Lipatov:1976zz,Kuraev:1976ge,Kuraev:1977fs,Balitsky:1978ic}.

\medskip
\section{Phenomenology and the early Universe with Higgsplosion}
\label{sec:pheno}

Many parts of models of the early Universe are relying on finite temperature effects and quantum fluctuations, all of which could receive corrections from the Higgsplosion mechanism, e.g. inflation \cite{Guth:1980zm,Linde:1981mu,Albrecht:1982wi,Linde:1983gd}, reheating \cite{Kofman:1997yn}, the cosmological microwave background \cite{Maldacena:2002vr,Crotty:2003rz}, blackhole formation \cite{Bonanno:1998ye, Koch:2013owa, Koch:2014cqa} and the vacuum energy density during the evolution of the Universe \cite{Ford:1992mv}. As these phenoma are not entirely independent and are deserving of a detailed investigation in their own right, a full study of the early history of the Universe in presence of Higgsplosion is beyond the scope of this work. However, in this section we briefly comment on whether and how radically Higgsplosion would change the standard Big Bang model of the early Universe.

\subsection{Higgsplosion and the running of gauge and gravity couplings}
\label{sec:gravity}
\medskip


General relativity is inherently difficult to reconcile with the quantum field theoretical description of the standard model. While the quantum theory of the standard model is predictive to all orders in perturbation theory, loop corrections to gravity can only be taken into account order by order and have to be treated in the context of an effective field theory with expansion parameter $E^2/M_{\mathrm{Pl}}^2$.
One way of addressing this problem is the concept of asymptotic safety \cite{Weinberg:1976xy,Weinberg:1980gg,Wetterich:1992yh,Reuter:1993kw,Reuter:1996cp,Litim:2008tt} which ensures that quantum field theories remain fundamental and predictive up to highest energies. This scenario indicates that a quantum theory of gravity can be renormalisable on a non-perturbative level, despite being perturbatively non-renormalisable. In gravity asymptotic safety aims to provide a path-integral framework where the metric field is the carrier of the fundamental degrees of freedom in the classical and quantum regime of the theory. Thus, the quantum field theoretical description of gravity can be extended to infinitely large energy scales. 
A realisation of asymptotic safety requires that the beta-functions of all couplings $g_i$ vanish at fix points $g_i^*$, i.e. $\beta(g_i^*) = 0$. The number of parameters $g_i$ defines the dimensionality of the ultraviolet critical surface formed by all trajectories attracted to the fixed point.

\begin{figure}[htp]
\begin{center}
\includegraphics[width=0.7\textwidth]{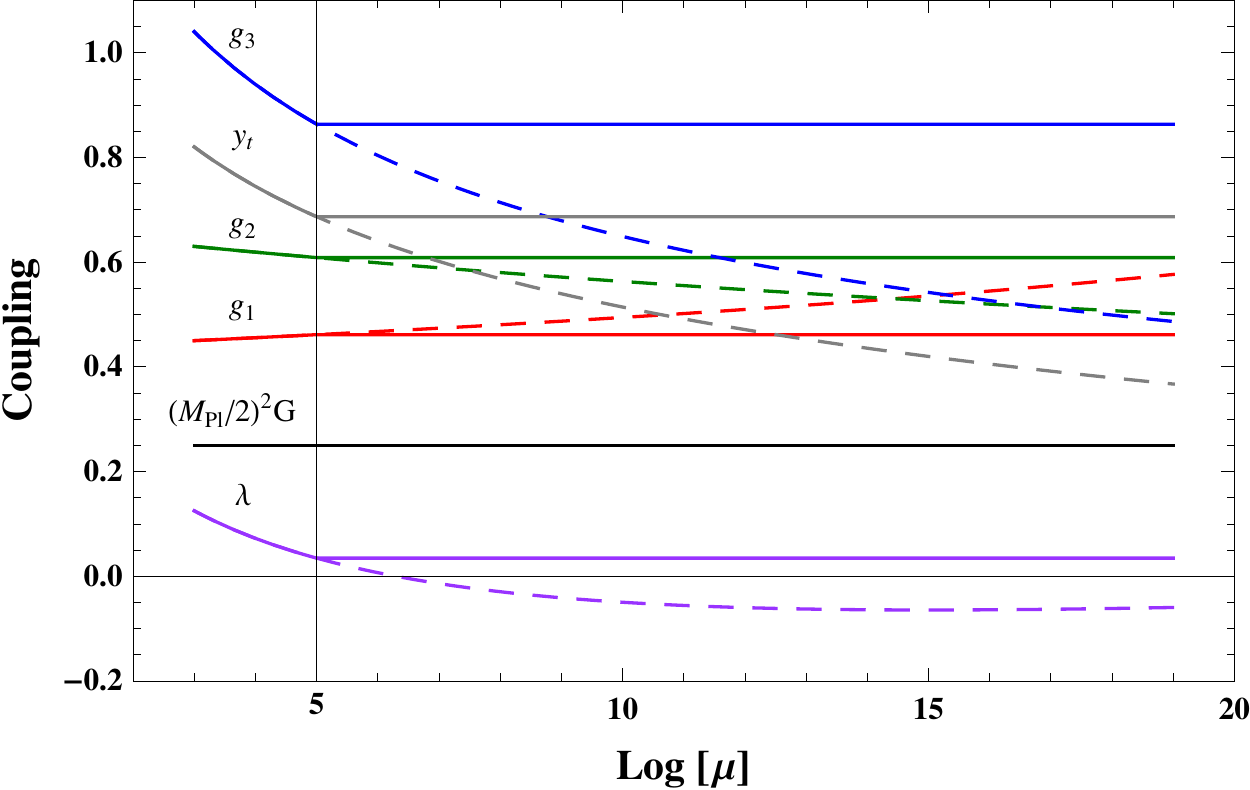}
\caption{Two-loop running of standard model couplings with Higgsplosion (solid lines) and without (dashed lines). We assumed the Higgsplosion scale to be uniformly at $E_*=10^5$ GeV.}
\label{fig:coupl}
\end{center}
\end{figure}

 In Fig.~\ref{fig:coupl} we show the impact of Higgsplosion on the running of the Yukawa, scalar, gauge and gravity couplings. For the standard model couplings we use 2-loop running as implemented in SARAH \cite{Staub:2008uz, Staub:2010jh}, while for gravity we show the classical value normalised to $(M_{\rm Pl}/2)^2$, as we do not want to speculate about the theory that governs quantum gravity in the UV. The couplings can have quantitatively different scales $\mu$ for $\beta(g_i)=0$, depending on the masses, couplings and stability of the degrees of freedom that drive their running. For simplicity we chose $E_*=10^5$ GeV for all couplings.

The whole SM exhibits a fix point with a finite-dimensional critical surface, as required for an asymptotically safe theory. Further, neither are there UV Landau poles associated with any of the gauge groups, Yukawa couplings or scalar interactions. In particular, if the Higgsplosion scale for all particles is below $\sim 10^6$ GeV, the Higgs potential remains stable on cosmological time scales.

While embedding the standard model into an asymptotically safe theory, i.e.~a theory free of Landau poles and free of a hierarchy problem due to Higgs-gravity interactions, has been a challenge \cite{Giudice:2014tma,Litim:2014uca,Abel:2017ujy}, within the Higgsplosion framework, following arguments of Sec.~\ref{sec:qft}, this is automatically realised for the minimal standard model and for most of its proposed extensions. In the standard model without Higgsplosion one expects gravity to give rise to a fine-tuning problem when $\delta M_h^2 \gtrsim M_h^2$, where $\delta M_h^2 \sim l G_N \Lambda_G^4$ with $l \sim (4 \pi)^{-4}$, i.e. around the scale $\Lambda_G \simeq 10^{11}$ GeV. However, with the graviton Higgsplosion scale $E_*^G$ being much smaller than $\Lambda_G$, as calculated in Sec.~\ref{sec:2.2.2}, the gravitational contributions to the fine-tuning of the Higgs mass is cut off at $\delta M_h^2 \sim l G_N (E_*^G)^4 \ll M_h^2$. Hence, in the Higgsplosion scenario, there is no introduction of a hierarchy problem due to gravity and the need to construct a scenario that softens gravity in the UV is absent.


\subsection{Higgsplosion during inflation}

In the standard Big Bang cosmology inflation was proposed to solve simultaneously the flatness, isotropy, homogenity, horizon and relics problems 
\cite{Guth:1980zm,Linde:1981mu,Albrecht:1982wi,Linde:1983gd}. Thus inflation is the most popular theory of the early universe, in full agreement with observations, including the recent data from Planck satellite \cite{Ade:2015lrj}, which favour a simple inflationary scenario with only one slow rolling scalar field. 

While an exhaustive discussion of inflation is beyond the scope of this article, we want to investigate if Higgsplosion can be reconciled with inflation or if one or the other has to be abandoned. We thus focus on a scenario with a non-minimally coupled scalar singlet field $S$ to gravity 
\cite{Salopek:1988qh} (see also \cite{Bezrukov:2007ep,Hertzberg:2010dc,Lebedev:2011aq,Khoze:2013uia})
as an example to show that Inflation can be incorporated into the Higgsplosion framework. A non-minimal inflaton coupling to gravity allows for a tensor-scalar ratio of $r_{0.05} \gtrsim -0.004$ well in agreement with current limits of $r_{0.05}<0.07$ \cite{Array:2015xqh}.
We take the relevant Jordan frame Lagrangian to be
\begin{equation}
\mathcal{L} = \sqrt{-g} \left [ -\frac{ M_{Pl}  + \xi_s S^2}{2}R + \partial_\mu H^\dagger \partial^\mu H + (\partial_\mu S)^2 - V(H,S)  \right ] 
\end{equation}
with the Higgs doublet $H = (\phi^+, 1/\sqrt{2} (h + i \phi^0))^T$. The inflaton's non-minimal coupling term to gravity $\xi_s S^2 R/2$ should have a large parameter $\xi_s \sim 10^4$. The tree-level two-field scalar potential is
\begin{equation}
V(H,S) = - \mu_h H^\dagger H + \lambda_h (H^\dagger H)^2 - \frac{1}{2} \mu_S^2 S^2 + \frac{1}{4} \lambda_S S^4 + \frac{1}{2} \lambda_{Sh} H^\dagger H S^2.
\end{equation}
To bound the potential from below we take all $\lambda_i$ to be positive.

The inflaton develops a vacuum expectation value $v_s$ during inflation. Thus, during inflation, the mass of both the inflaton and the Higgs boson in the inflaton background are large
\begin{equation}
M_h \simeq \sqrt{ \frac{ \lambda_{Sh}} {2} } S(x) \simeq \frac{M_{Pl}}{\sqrt{\xi_s}}.
\end{equation}

As $h$ is of the order of the mass of the inflation $S$, or even heavier, the inflaton cannot higgsplode during inflation. Phenomenologically, inflation within the singlet extended Standard Model can remain unaffected by Higgsplosion.

The picture can change during reheating, where the inflaton oscillates around the potential's minimum, and with it the Higgs mass varies. One could imagine that reheating becomes more efficient if Higgsplosion sets in and that the mechanism is different from standard resonant reheating, which is associated with growing classical instabilities. A detailed study of reheating seems warranted but is beyond the scope of this work.

\subsection{Axions}

The strong CP problem, the discrepancy between the theoretically allowed value of the sum of the QCD topological angle and the quark mass phase $\theta = \theta_0 + \arg \det M_q$ and its experimentally observed size of less than $\mathcal{O}(10^{-10})$, 
provides a motivation to augment the SM by an additional pseudo-nambu-goldstone boson $a$ of a spontaneously broken $U(1)_{PQ}$ symmetry \cite{Peccei:1977ur,Peccei:1977hh,Weinberg:1977ma,Wilczek:1977pj,Kim:1979if,Shifman:1979if,Dine:1981rt}. The axion's Lagrangian below the Peccei Quinn (PQ) breaking scale can be written as 
\begin{equation}
\mathcal{L}_a = \frac{1}{2}(\partial_\mu a)^2 + \frac{a}{f_a}\frac{\alpha_s}{8 \pi} G_{\mu\nu} \tilde{G}^{\mu \nu} + \frac{a}{4} g^{0}_{a \gamma \gamma} F_{\mu \nu} \tilde{F}^{\mu \nu} + \frac{\partial_\mu a}{2 f_a} g_q \bar{q} \gamma^{\mu} \gamma_5 q
\end{equation}
where the axion decay constant $f_a$ is the order parameter associated with the breaking of $U(1)_{PQ}$ via $a(x) \to a(x) + \alpha f_a$ and the dual gluon field strength $\tilde{G}_{\mu \nu} = \frac{1}{2} \epsilon_{\mu \nu \rho \sigma} G^{\rho \sigma}$ (analogously for $\tilde{F}^{\mu \nu}$).

In the context of Higgsplosion, the QCD-axion provides a well-defined framework to answer the question if light degrees of freedom are allowed, or if they would lower the Higgsplosion scale enough to render this mechanism incompatible with the existence of very light and weakly coupled scalars. The only free parameter that is defining the axion's interactions and mass is $f_a$. At next-to-leading order\footnote{While it is conceptually inconsistent to use a next-to-leading order result calculated without Higgsplosion, we do not expect a large quantitative effects from discarding the high loop momenta in the calculation of $m_a$ and $\lambda_a$.} the axion mass and self interaction  are respectively calculated to be \cite{diCortona:2015ldu}
\begin{equation}
m_a \simeq  \frac{5.7\cdot 10^{15}\mathrm{eV}}{f_a} 
\end{equation}
and
\begin{equation}
\lambda_a \equiv \left . \frac{\partial^4 V(a)}{\partial a^4} \right |_{a=0} \simeq -0.346 \frac{m^2_a}{f^2_a}.
\end{equation}

We can now estimate the Higgsplosion scale where $\alpha_s$ stops running due to axion contributions in the Higgsplosion of the gluon. A crude approximation, assuming $\lambda_a n_a \gtrsim 20$ results in Higgsplosion, gives a Higgsplosion scale $E_*$ of 
\begin{equation}
E^{\mathrm{Axion}}_* \simeq 60 \frac{f_a^2}{m_a}.
\end{equation}
If we require the Axion's Higgsplosion\footnote{To avoid confusion we will refrain from calling the rapid increase of the transition amplitude in the process $g \to g~n \times a$ Axionplosion, and will just refer to it as the Axion's Higgsplosion process.} scale to be above the Higgs boson's Higgsplosion scale, i.e. $E^{\mathrm{Axion}}_* > 10^5$ GeV, we find a limit $f_a\gtrsim 2.1$ GeV, which is easily achievable. Such a bound is far below existing experimental limits of $f_a\gtrsim 10^8-10^{17}$ GeV \cite{Raffelt:2006cw,Arvanitaki:2009fg,Arvanitaki:2010sy,Arvanitaki:2014wva}. 

Thus, the existence of QCD axions is not in conflict with Higgsplosion and the axion's  contribution to the Higgsplosion of SM light degrees of freedom is negligible for sufficiently large $f_a$.

\medskip
\section{Discussion and Conclusions}
\label{sec:concl}

The discovery of the Higgs boson has unravelled an extraordinary building block in our understanding of elementary particle physics: the first elementary scalar particle. An immense effort is being devoted to determining its precise properties and in using it as a vehicle to uncover answers to thus far inexplicable observations in nature. 
Yet, the Higgs boson's hierarchy problem and its peculiar contribution to $h^* \to n \times h$ transition amplitudes have been puzzling for a long time and have provided motivation to extend the standard model of particle physics by novel degrees of freedom and interactions. In \cite{Khoze:2017tjt} we have proposed two intertwined mechanisms, Higgsplosion and Higgspersion, to address both of these issues. In this paper work we have extended the discussion and interpretation of Higgsplosion and Higgspersion and have outlined some phenomenological, i.e. experimentally testable, consequences in case these mechanisms are realised in nature.

Our findings are:

\begin{enumerate}
\item All particles of the standard model and its extensions higgsplode at individually different scales. The particle's Higgsplosion scale depends on its interaction strength with the Higgs boson, its mass and whether it is stable or not.
\item The Higgsplosion scale establishes a radius $E_* \sim 1/r_*$ that cannot be probed further inside by increasing momenta. Thus, interaction processes with propagators with $p^2 \geq E_*^2$, i.e. all $s$-channel processes shut down, pronouncing the importance of $t$-channel interactions, e.g. multi-regge kineamtics, in high-energy scatterings. $E_*$ effectively defines a virtuality and interaction dependent compositeness or classicalization scale.
\item Not only tree-level interactions are higgspersed, but $n$-point functions to all loop orders are cut-off as well at $p^2=E^2_*$. Hence, beyond the Higgsplosion scale the beta functions of all couplings vanish, i.e. $\beta(g_i^*)=0$, and the standard model becomes asymptotically safe and free of Landau poles.
\item If additional scalars are in the theory that are either heavy, have a small mass ratio to the Higgs boson, or are feeble coupled they will not higgsplode or provide a significant contribution to the Higgsplosion of other particles. Thus, inflation and the existence of QCD axions are not in conflict with the Higgsplosion mechanism.
\end{enumerate}

\bigskip

\noindent
Arguably the effects of Higgsplosion and Higgspersion open up to a radical re-examination of the standard wisdom associated with the
UV behaviour in quantum field theory, its phenomenological applications, and its probes of nature.
Many fundamental questions are as yet left unanswered and we believe that this offers potential for exciting discoveries and warrant further studies in this area.


\bigskip

\section*{Acknowledgements}
VVK would like to thank our colleagues at Tel Aviv University and the Weizmann Institute for hospitality and stimulating discussions while part of this work was completed. 
We thank Kfir Blum, Gia Dvali, Joerg Jaeckel, Marek Karliner, Daniel Litim, Gilad Perez, Carlos Tamarit, Lorenzo Ubaldi and Tomer Volansky for helpful discussions.


\bibliographystyle{JHEP}
\bibliography{references}

\providecommand{\href}[2]{#2}\begingroup\raggedright\begin{thebibliography}{10}

\bibitem{Khoze:2017tjt}
V.~V. Khoze and M.~Spannowsky, \emph{{Higgsplosion: Solving the Hierarchy
  Problem via rapid decays of heavy states into multiple Higgs bosons}},
  \href{https://arxiv.org/abs/1704.03447}{{\tt 1704.03447}}.

\bibitem{Brown:1992ay}
L.~S. Brown, \emph{{Summing tree graphs at threshold}},
  \href{http://dx.doi.org/10.1103/PhysRevD.46.R4125}{\emph{Phys. Rev.} {\bf
  D46} (1992) R4125--R4127}, [\href{https://arxiv.org/abs/hep-ph/9209203}{{\tt
  hep-ph/9209203}}].

\bibitem{Argyres:1992np}
E.~N. Argyres, R.~H.~P. Kleiss and C.~G. Papadopoulos, \emph{{Amplitude
  estimates for multi - Higgs production at high-energies}},
  \href{http://dx.doi.org/10.1016/0550-3213(93)90140-K}{\emph{Nucl. Phys.} {\bf
  B391} (1993) 42--56}.

\bibitem{Voloshin:1992rr}
M.~B. Voloshin, \emph{{Estimate of the onset of nonperturbative particle
  production at high-energy in a scalar theory}},
  \href{http://dx.doi.org/10.1016/0370-2693(92)90901-F}{\emph{Phys. Lett.} {\bf
  B293} (1992) 389--394}.

\bibitem{Libanov:1994ug}
M.~V. Libanov, V.~A. Rubakov, D.~T. Son and S.~V. Troitsky,
  \emph{{Exponentiation of multiparticle amplitudes in scalar theories}},
  \href{http://dx.doi.org/10.1103/PhysRevD.50.7553}{\emph{Phys. Rev.} {\bf D50}
  (1994) 7553--7569}, [\href{https://arxiv.org/abs/hep-ph/9407381}{{\tt
  hep-ph/9407381}}].

\bibitem{Khoze:2014kka}
V.~V. Khoze, \emph{{Perturbative growth of high-multiplicity W, Z and Higgs
  production processes at high energies}},
  \href{http://dx.doi.org/10.1007/JHEP03(2015)038}{\emph{JHEP} {\bf 03} (2015)
  038}, [\href{https://arxiv.org/abs/1411.2925}{{\tt 1411.2925}}].

\bibitem{Voloshin:1992nu}
M.~B. Voloshin, \emph{{Summing one loop graphs at multiparticle threshold}},
  \href{http://dx.doi.org/10.1103/PhysRevD.47.R357}{\emph{Phys. Rev.} {\bf D47}
  (1993) R357--R361}, [\href{https://arxiv.org/abs/hep-ph/9209240}{{\tt
  hep-ph/9209240}}].

\bibitem{Smith:1992rq}
B.~H. Smith, \emph{{Summing one loop graphs in a theory with broken symmetry}},
  \href{http://dx.doi.org/10.1103/PhysRevD.47.3518}{\emph{Phys. Rev.} {\bf D47}
  (1993) 3518--3520}, [\href{https://arxiv.org/abs/hep-ph/9209287}{{\tt
  hep-ph/9209287}}].

\bibitem{Son:1995wz}
D.~T. Son, \emph{{Semiclassical approach for multiparticle production in scalar
  theories}}, \href{http://dx.doi.org/10.1016/0550-3213(96)00386-0}{\emph{Nucl.
  Phys.} {\bf B477} (1996) 378--406},
  [\href{https://arxiv.org/abs/hep-ph/9505338}{{\tt hep-ph/9505338}}].

\bibitem{Gorsky:1993ix}
A.~S. Gorsky and M.~B. Voloshin, \emph{{Nonperturbative production of
  multiboson states and quantum bubbles}},
  \href{http://dx.doi.org/10.1103/PhysRevD.48.3843}{\emph{Phys. Rev.} {\bf D48}
  (1993) 3843--3851}, [\href{https://arxiv.org/abs/hep-ph/9305219}{{\tt
  hep-ph/9305219}}].

\bibitem{Khoze:2017ifq}
V.~V. Khoze, \emph{{Multiparticle production in the large lambda n limit:
  Realising Higgsplosion in a scalar QFT}},
  \href{https://arxiv.org/abs/1705.04365}{{\tt 1705.04365}}.

\bibitem{Aad:2012tfa}
{\scshape ATLAS} collaboration, G.~Aad et~al., \emph{{Observation of a new
  particle in the search for the Standard Model Higgs boson with the ATLAS
  detector at the LHC}},
  \href{http://dx.doi.org/10.1016/j.physletb.2012.08.020}{\emph{Phys. Lett.}
  {\bf B716} (2012) 1--29}, [\href{https://arxiv.org/abs/1207.7214}{{\tt
  1207.7214}}].

\bibitem{Chatrchyan:2012xdj}
{\scshape CMS} collaboration, S.~Chatrchyan et~al., \emph{{Observation of a new
  boson at a mass of 125 GeV with the CMS experiment at the LHC}},
  \href{http://dx.doi.org/10.1016/j.physletb.2012.08.021}{\emph{Phys. Lett.}
  {\bf B716} (2012) 30--61}, [\href{https://arxiv.org/abs/1207.7235}{{\tt
  1207.7235}}].

\bibitem{Contino:2016spe}
R.~Contino et~al., \emph{{Physics at a 100 TeV pp collider: Higgs and EW
  symmetry breaking studies}},
  \href{http://dx.doi.org/10.23731/CYRM-2017-003.255}{\emph{CERN Yellow Report}
  (2017) 255--440}, [\href{https://arxiv.org/abs/1606.09408}{{\tt
  1606.09408}}].

\bibitem{Voloshin:1994yp}
M.~B. Voloshin, \emph{{Nonperturbative methods}},  in \emph{{27th International
  Conference on High-energy Physics (ICHEP 94) Glasgow, Scotland, July 20-27,
  1994}}, pp.~0121--134, 1994.
\newblock \href{https://arxiv.org/abs/hep-ph/9409344}{{\tt hep-ph/9409344}}.

\bibitem{Libanov:1997nt}
M.~V. Libanov, V.~A. Rubakov and S.~V. Troitsky, \emph{{Multiparticle processes
  and semiclassical analysis in bosonic field theories}},
  \href{http://dx.doi.org/10.1134/1.953038}{\emph{Phys. Part. Nucl.} {\bf 28}
  (1997) 217--240}.

\bibitem{Khoze:2014zha}
V.~V. Khoze, \emph{{Multiparticle Higgs and Vector Boson Amplitudes at
  Threshold}}, \href{http://dx.doi.org/10.1007/JHEP07(2014)008}{\emph{JHEP}
  {\bf 07} (2014) 008}, [\href{https://arxiv.org/abs/1404.4876}{{\tt
  1404.4876}}].

\bibitem{Donoghue:2017pgk}
J.~F. Donoghue, M.~M. Ivanov and A.~Shkerin, \emph{{EPFL Lectures on General
  Relativity as a Quantum Field Theory}},
  \href{https://arxiv.org/abs/1702.00319}{{\tt 1702.00319}}.

\bibitem{Wilson:1971bg}
K.~G. Wilson, \emph{{Renormalization group and critical phenomena. 1.
  Renormalization group and the Kadanoff scaling picture}},
  \href{http://dx.doi.org/10.1103/PhysRevB.4.3174}{\emph{Phys. Rev.} {\bf B4}
  (1971) 3174--3183}.

\bibitem{Wilson:1973jj}
K.~G. Wilson and J.~B. Kogut, \emph{{The Renormalization group and the epsilon
  expansion}},
  \href{http://dx.doi.org/10.1016/0370-1573(74)90023-4}{\emph{Phys. Rept.} {\bf
  12} (1974) 75--200}.

\bibitem{Polchinski:1983gv}
J.~Polchinski, \emph{{Renormalization and Effective Lagrangians}},
  \href{http://dx.doi.org/10.1016/0550-3213(84)90287-6}{\emph{Nucl. Phys.} {\bf
  B231} (1984) 269--295}.

\bibitem{Dvali:2010jz}
G.~Dvali, G.~F. Giudice, C.~Gomez and A.~Kehagias, \emph{{UV-Completion by
  Classicalization}},
  \href{http://dx.doi.org/10.1007/JHEP08(2011)108}{\emph{JHEP} {\bf 08} (2011)
  108}, [\href{https://arxiv.org/abs/1010.1415}{{\tt 1010.1415}}].

\bibitem{Dvali:2016ovn}
G.~Dvali, \emph{{Strong Coupling and Classicalization}},  in
  \emph{{Proceedings, LHCSki 2016 - A First Discussion of 13 TeV Results:
  Obergurgl, Austria, April 10-15, 2016}}, 2016.
\newblock \href{https://arxiv.org/abs/1607.07422}{{\tt 1607.07422}}.
\newblock \href{http://dx.doi.org/10.1142/9789813208292_0005}{DOI}.

\bibitem{Lipatov:1976zz}
L.~N. Lipatov, \emph{{Reggeization of the Vector Meson and the Vacuum
  Singularity in Nonabelian Gauge Theories}}, {\emph{Sov. J. Nucl. Phys.} {\bf
  23} (1976) 338--345}.

\bibitem{Kuraev:1976ge}
E.~A. Kuraev, L.~N. Lipatov and V.~S. Fadin, \emph{{Multi - Reggeon Processes
  in the Yang-Mills Theory}}, {\emph{Sov. Phys. JETP} {\bf 44} (1976)
  443--450}.

\bibitem{Kuraev:1977fs}
E.~A. Kuraev, L.~N. Lipatov and V.~S. Fadin, \emph{{The Pomeranchuk Singularity
  in Nonabelian Gauge Theories}}, {\emph{Sov. Phys. JETP} {\bf 45} (1977)
  199--204}.

\bibitem{Balitsky:1978ic}
I.~I. Balitsky and L.~N. Lipatov, \emph{{The Pomeranchuk Singularity in Quantum
  Chromodynamics}}, {\emph{Sov. J. Nucl. Phys.} {\bf 28} (1978) 822--829}.

\bibitem{Guth:1980zm}
A.~H. Guth, \emph{{The Inflationary Universe: A Possible Solution to the
  Horizon and Flatness Problems}},
  \href{http://dx.doi.org/10.1103/PhysRevD.23.347}{\emph{Phys. Rev.} {\bf D23}
  (1981) 347--356}.

\bibitem{Linde:1981mu}
A.~D. Linde, \emph{{A New Inflationary Universe Scenario: A Possible Solution
  of the Horizon, Flatness, Homogeneity, Isotropy and Primordial Monopole
  Problems}}, \href{http://dx.doi.org/10.1016/0370-2693(82)91219-9}{\emph{Phys.
  Lett.} {\bf 108B} (1982) 389--393}.

\bibitem{Albrecht:1982wi}
A.~Albrecht and P.~J. Steinhardt, \emph{{Cosmology for Grand Unified Theories
  with Radiatively Induced Symmetry Breaking}},
  \href{http://dx.doi.org/10.1103/PhysRevLett.48.1220}{\emph{Phys. Rev. Lett.}
  {\bf 48} (1982) 1220--1223}.

\bibitem{Linde:1983gd}
A.~D. Linde, \emph{{Chaotic Inflation}},
  \href{http://dx.doi.org/10.1016/0370-2693(83)90837-7}{\emph{Phys. Lett.} {\bf
  129B} (1983) 177--181}.

\bibitem{Kofman:1997yn}
L.~Kofman, A.~D. Linde and A.~A. Starobinsky, \emph{{Towards the theory of
  reheating after inflation}},
  \href{http://dx.doi.org/10.1103/PhysRevD.56.3258}{\emph{Phys. Rev.} {\bf D56}
  (1997) 3258--3295}, [\href{https://arxiv.org/abs/hep-ph/9704452}{{\tt
  hep-ph/9704452}}].

\bibitem{Maldacena:2002vr}
J.~M. Maldacena, \emph{{Non-Gaussian features of primordial fluctuations in
  single field inflationary models}},
  \href{http://dx.doi.org/10.1088/1126-6708/2003/05/013}{\emph{JHEP} {\bf 05}
  (2003) 013}, [\href{https://arxiv.org/abs/astro-ph/0210603}{{\tt
  astro-ph/0210603}}].

\bibitem{Crotty:2003rz}
P.~Crotty, J.~Garcia-Bellido, J.~Lesgourgues and A.~Riazuelo, \emph{{Bounds on
  isocurvature perturbations from CMB and LSS data}},
  \href{http://dx.doi.org/10.1103/PhysRevLett.91.171301}{\emph{Phys. Rev.
  Lett.} {\bf 91} (2003) 171301},
  [\href{https://arxiv.org/abs/astro-ph/0306286}{{\tt astro-ph/0306286}}].

\bibitem{Bonanno:1998ye}
A.~Bonanno and M.~Reuter, \emph{{Quantum gravity effects near the null black
  hole singularity}},
  \href{http://dx.doi.org/10.1103/PhysRevD.60.084011}{\emph{Phys. Rev.} {\bf
  D60} (1999) 084011}, [\href{https://arxiv.org/abs/gr-qc/9811026}{{\tt
  gr-qc/9811026}}].

\bibitem{Koch:2013owa}
B.~Koch and F.~Saueressig, \emph{{Structural aspects of asymptotically safe
  black holes}},
  \href{http://dx.doi.org/10.1088/0264-9381/31/1/015006}{\emph{Class. Quant.
  Grav.} {\bf 31} (2014) 015006}, [\href{https://arxiv.org/abs/1306.1546}{{\tt
  1306.1546}}].

\bibitem{Koch:2014cqa}
B.~Koch and F.~Saueressig, \emph{{Black holes within Asymptotic Safety}},
  \href{http://dx.doi.org/10.1142/S0217751X14300117}{\emph{Int. J. Mod. Phys.}
  {\bf A29} (2014) 1430011}, [\href{https://arxiv.org/abs/1401.4452}{{\tt
  1401.4452}}].

\bibitem{Ford:1992mv}
C.~Ford, D.~R.~T. Jones, P.~W. Stephenson and M.~B. Einhorn, \emph{{The
  Effective potential and the renormalization group}},
  \href{http://dx.doi.org/10.1016/0550-3213(93)90206-5}{\emph{Nucl. Phys.} {\bf
  B395} (1993) 17--34}, [\href{https://arxiv.org/abs/hep-lat/9210033}{{\tt
  hep-lat/9210033}}].

\bibitem{Weinberg:1976xy}
S.~Weinberg, \emph{{Critical Phenomena for Field Theorists}},  in \emph{{14th
  International School of Subnuclear Physics: Understanding the Fundamental
  Constitutents of Matter Erice, Italy, July 23-August 8, 1976}}, p.~1, 1976.

\bibitem{Weinberg:1980gg}
S.~Weinberg, \emph{{ULTRAVIOLET DIVERGENCES IN QUANTUM THEORIES OF
  GRAVITATION}},  in \emph{General Relativity: An Einstein Centenary Survey},
  pp.~790--831.
\newblock 1980.

\bibitem{Wetterich:1992yh}
C.~Wetterich, \emph{{Exact evolution equation for the effective potential}},
  \href{http://dx.doi.org/10.1016/0370-2693(93)90726-X}{\emph{Phys. Lett.} {\bf
  B301} (1993) 90--94}.

\bibitem{Reuter:1993kw}
M.~Reuter and C.~Wetterich, \emph{{Effective average action for gauge theories
  and exact evolution equations}},
  \href{http://dx.doi.org/10.1016/0550-3213(94)90543-6}{\emph{Nucl. Phys.} {\bf
  B417} (1994) 181--214}.

\bibitem{Reuter:1996cp}
M.~Reuter, \emph{{Nonperturbative evolution equation for quantum gravity}},
  \href{http://dx.doi.org/10.1103/PhysRevD.57.971}{\emph{Phys. Rev.} {\bf D57}
  (1998) 971--985}, [\href{https://arxiv.org/abs/hep-th/9605030}{{\tt
  hep-th/9605030}}].

\bibitem{Litim:2008tt}
D.~F. Litim, \emph{{Fixed Points of Quantum Gravity and the Renormalisation
  Group}},  \href{https://arxiv.org/abs/0810.3675}{{\tt 0810.3675}}.

\bibitem{Staub:2008uz}
F.~Staub, \emph{{SARAH}},  \href{https://arxiv.org/abs/0806.0538}{{\tt
  0806.0538}}.

\bibitem{Staub:2010jh}
F.~Staub, \emph{{Automatic Calculation of supersymmetric Renormalization Group
  Equations and Self Energies}},
  \href{http://dx.doi.org/10.1016/j.cpc.2010.11.030}{\emph{Comput. Phys.
  Commun.} {\bf 182} (2011) 808--833},
  [\href{https://arxiv.org/abs/1002.0840}{{\tt 1002.0840}}].

\bibitem{Giudice:2014tma}
G.~F. Giudice, G.~Isidori, A.~Salvio and A.~Strumia, \emph{{Softened Gravity
  and the Extension of the Standard Model up to Infinite Energy}},
  \href{http://dx.doi.org/10.1007/JHEP02(2015)137}{\emph{JHEP} {\bf 02} (2015)
  137}, [\href{https://arxiv.org/abs/1412.2769}{{\tt 1412.2769}}].

\bibitem{Litim:2014uca}
D.~F. Litim and F.~Sannino, \emph{{Asymptotic safety guaranteed}},
  \href{http://dx.doi.org/10.1007/JHEP12(2014)178}{\emph{JHEP} {\bf 12} (2014)
  178}, [\href{https://arxiv.org/abs/1406.2337}{{\tt 1406.2337}}].

\bibitem{Abel:2017ujy}
S.~Abel and F.~Sannino, \emph{{Radiative symmetry breaking from interacting UV
  fixed points}},  \href{https://arxiv.org/abs/1704.00700}{{\tt 1704.00700}}.

\bibitem{Ade:2015lrj}
{\scshape Planck} collaboration, P.~A.~R. Ade et~al., \emph{{Planck 2015
  results. XX. Constraints on inflation}},
  \href{http://dx.doi.org/10.1051/0004-6361/201525898}{\emph{Astron.
  Astrophys.} {\bf 594} (2016) A20},
  [\href{https://arxiv.org/abs/1502.02114}{{\tt 1502.02114}}].

\bibitem{Salopek:1988qh}
D.~S. Salopek, J.~R. Bond and J.~M. Bardeen, \emph{{Designing Density
  Fluctuation Spectra in Inflation}},
  \href{http://dx.doi.org/10.1103/PhysRevD.40.1753}{\emph{Phys. Rev.} {\bf D40}
  (1989) 1753}.

\bibitem{Bezrukov:2007ep}
F.~L. Bezrukov and M.~Shaposhnikov, \emph{{The Standard Model Higgs boson as
  the inflaton}},
  \href{http://dx.doi.org/10.1016/j.physletb.2007.11.072}{\emph{Phys. Lett.}
  {\bf B659} (2008) 703--706}, [\href{https://arxiv.org/abs/0710.3755}{{\tt
  0710.3755}}].

\bibitem{Hertzberg:2010dc}
M.~P. Hertzberg, \emph{{On Inflation with Non-minimal Coupling}},
  \href{http://dx.doi.org/10.1007/JHEP11(2010)023}{\emph{JHEP} {\bf 11} (2010)
  023}, [\href{https://arxiv.org/abs/1002.2995}{{\tt 1002.2995}}].

\bibitem{Lebedev:2011aq}
O.~Lebedev and H.~M. Lee, \emph{{Higgs Portal Inflation}},
  \href{http://dx.doi.org/10.1140/epjc/s10052-011-1821-0}{\emph{Eur. Phys. J.}
  {\bf C71} (2011) 1821}, [\href{https://arxiv.org/abs/1105.2284}{{\tt
  1105.2284}}].

\bibitem{Khoze:2013uia}
V.~V. Khoze, \emph{{Inflation and Dark Matter in the Higgs Portal of
  Classically Scale Invariant Standard Model}},
  \href{http://dx.doi.org/10.1007/JHEP11(2013)215}{\emph{JHEP} {\bf 11} (2013)
  215}, [\href{https://arxiv.org/abs/1308.6338}{{\tt 1308.6338}}].

\bibitem{Array:2015xqh}
{\scshape BICEP2, Keck Array} collaboration, P.~A.~R. Ade et~al.,
  \emph{{Improved Constraints on Cosmology and Foregrounds from BICEP2 and Keck
  Array Cosmic Microwave Background Data with Inclusion of 95 GHz Band}},
  \href{http://dx.doi.org/10.1103/PhysRevLett.116.031302}{\emph{Phys. Rev.
  Lett.} {\bf 116} (2016) 031302},
  [\href{https://arxiv.org/abs/1510.09217}{{\tt 1510.09217}}].

\bibitem{Peccei:1977ur}
R.~D. Peccei and H.~R. Quinn, \emph{{Constraints Imposed by CP Conservation in
  the Presence of Instantons}},
  \href{http://dx.doi.org/10.1103/PhysRevD.16.1791}{\emph{Phys. Rev.} {\bf D16}
  (1977) 1791--1797}.

\bibitem{Peccei:1977hh}
R.~D. Peccei and H.~R. Quinn, \emph{{CP Conservation in the Presence of
  Instantons}},
  \href{http://dx.doi.org/10.1103/PhysRevLett.38.1440}{\emph{Phys. Rev. Lett.}
  {\bf 38} (1977) 1440--1443}.

\bibitem{Weinberg:1977ma}
S.~Weinberg, \emph{{A New Light Boson?}},
  \href{http://dx.doi.org/10.1103/PhysRevLett.40.223}{\emph{Phys. Rev. Lett.}
  {\bf 40} (1978) 223--226}.

\bibitem{Wilczek:1977pj}
F.~Wilczek, \emph{{Problem of Strong p and t Invariance in the Presence of
  Instantons}}, \href{http://dx.doi.org/10.1103/PhysRevLett.40.279}{\emph{Phys.
  Rev. Lett.} {\bf 40} (1978) 279--282}.

\bibitem{Kim:1979if}
J.~E. Kim, \emph{{Weak Interaction Singlet and Strong CP Invariance}},
  \href{http://dx.doi.org/10.1103/PhysRevLett.43.103}{\emph{Phys. Rev. Lett.}
  {\bf 43} (1979) 103}.

\bibitem{Shifman:1979if}
M.~A. Shifman, A.~I. Vainshtein and V.~I. Zakharov, \emph{{Can Confinement
  Ensure Natural CP Invariance of Strong Interactions?}},
  \href{http://dx.doi.org/10.1016/0550-3213(80)90209-6}{\emph{Nucl. Phys.} {\bf
  B166} (1980) 493--506}.

\bibitem{Dine:1981rt}
M.~Dine, W.~Fischler and M.~Srednicki, \emph{{A Simple Solution to the Strong
  CP Problem with a Harmless Axion}},
  \href{http://dx.doi.org/10.1016/0370-2693(81)90590-6}{\emph{Phys. Lett.} {\bf
  104B} (1981) 199--202}.

\bibitem{diCortona:2015ldu}
G.~Grilli~di Cortona, E.~Hardy, J.~Pardo~Vega and G.~Villadoro, \emph{{The QCD
  axion, precisely}},
  \href{http://dx.doi.org/10.1007/JHEP01(2016)034}{\emph{JHEP} {\bf 01} (2016)
  034}, [\href{https://arxiv.org/abs/1511.02867}{{\tt 1511.02867}}].

\bibitem{Raffelt:2006cw}
G.~G. Raffelt, \emph{{Astrophysical axion bounds}},
  \href{http://dx.doi.org/10.1007/978-3-540-73518-2_3}{\emph{Lect. Notes Phys.}
  {\bf 741} (2008) 51--71}, [\href{https://arxiv.org/abs/hep-ph/0611350}{{\tt
  hep-ph/0611350}}].

\bibitem{Arvanitaki:2009fg}
A.~Arvanitaki, S.~Dimopoulos, S.~Dubovsky, N.~Kaloper and J.~March-Russell,
  \emph{{String Axiverse}},
  \href{http://dx.doi.org/10.1103/PhysRevD.81.123530}{\emph{Phys. Rev.} {\bf
  D81} (2010) 123530}, [\href{https://arxiv.org/abs/0905.4720}{{\tt
  0905.4720}}].

\bibitem{Arvanitaki:2010sy}
A.~Arvanitaki and S.~Dubovsky, \emph{{Exploring the String Axiverse with
  Precision Black Hole Physics}},
  \href{http://dx.doi.org/10.1103/PhysRevD.83.044026}{\emph{Phys. Rev.} {\bf
  D83} (2011) 044026}, [\href{https://arxiv.org/abs/1004.3558}{{\tt
  1004.3558}}].

\bibitem{Arvanitaki:2014wva}
A.~Arvanitaki, M.~Baryakhtar and X.~Huang, \emph{{Discovering the QCD Axion
  with Black Holes and Gravitational Waves}},
  \href{http://dx.doi.org/10.1103/PhysRevD.91.084011}{\emph{Phys. Rev.} {\bf
  D91} (2015) 084011}, [\href{https://arxiv.org/abs/1411.2263}{{\tt
  1411.2263}}].

\end{thebibliography}\endgroup

\end{document}